\documentclass[twocolumn,aps,prd,showpacs,nofootinbib,superscriptaddress,preprintnumbers,floatfix,10pt]{revtex4-1}

\usepackage[english]{babel} 
\usepackage[utf8]{inputenc}
\usepackage[T1]{fontenc}
\usepackage{lmodern}

\usepackage{amsmath}
\usepackage{amsfonts}
\usepackage{amssymb}
\usepackage{slashed}
\usepackage{bbm}

\usepackage{graphicx}
\usepackage{color}
\usepackage{multirow}

\usepackage{hyperref}

\newcommand{\I}{\mathrm{i}}
\newcommand{\E}{\mathrm{e}}
\newcommand{\tr}{\mathrm{tr}}

\newcommand{\diag}{\mathrm{diag}}

\newcommand{\pd}{\partial}
\newcommand{\pdm}{\pd_{\mu}}

\newcommand{\im}{\mathrm{Im}}
\newcommand{\re}{\mathrm{Re}}
\newcommand{\mn}{{\mu\nu}}
\newcommand{\m}{{\mu}}

\newcommand{\SUN}{\mathrm{SU}(N)}
\newcommand{\SU}{\mathrm{SU}}
\newcommand{\U}{\mathrm{U}}

\newcommand{\mh}{m_{\mathrm{h}}}
\newcommand{\mH}{m_{\mathrm{H}}}
\newcommand{\mA}{m_{\mathrm{A}}}
\newcommand{\MA}{M_{\mathrm{A}}}

\newcommand{\bma}{\begin{pmatrix}}
\newcommand{\ema}{\end{pmatrix}}

\definecolor{pt}{rgb}{0.0, 0.5, 0.0}

\begin{document}

\title{On the observable spectrum of theories with a Brout-Englert-Higgs effect}

\author{A.~Maas}
\email{axel.maas@uni-graz.at}
\affiliation{Institute of Physics, NAWI Graz, University of Graz, Universit\"atsplatz 5, 8010 Graz, Austria}

\author{R.~Sondenheimer}
\email{rene.sondenheimer@uni-jena.de}
\affiliation{Theoretisch-Physikalisches Institut, Friedrich-Schiller-Universit\"at Jena, Max-Wien-Platz 1, 07743 Jena, Germany}

\author{P.~T\"orek}
\email{pascal.toerek@uni-graz.at}
\affiliation{Institute of Physics, NAWI Graz, University of Graz, Universit\"atsplatz 5, 8010 Graz, Austria}

\begin{abstract}
The physical, observable spectrum in gauge theories is made up from gauge-invariant states. The Fr\"ohlich-Morchio-Strocchi mechanism allows in the standard model to map these states to the gauge-dependent elementary $W$, $Z$ and Higgs states. This is no longer necessarily the case in theories with a more general gauge group and Higgs sector. We classify and predict the physical spectrum for a wide range of such theories, with special emphasis on GUT-like cases, and show that discrepancies between the spectrum of elementary fields and physical particles frequently arise.
\end{abstract}

\pacs{}
\maketitle

\section{Introduction}

The physical states of gauge theories need to be manifestly and non-perturbatively gauge-invariant. This is realized in different ways in the different sectors of the standard model. In QCD confinement leads to only gauge-invariant states in the spectrum \cite{Weinberg:1996kr}. In QED dressings with Dirac phases create physical states \cite{Haag:1992hx,Lavelle:1995ty}. But also in the weak sector a non-trivial, non-perturbative mechanism turns out to be necessary \cite{'tHooft:1979bj,Osterwalder:1977pc,Banks:1979fi,Frohlich:1980gj}, the Fr\"ohlich-Morchio-Strocchi (FMS) mechanism to be discussed in detail in Sec.~\ref{sec:recipie} \cite{Frohlich:1980gj,Frohlich:1981yi}.

At first sight, this seems surprising. Standard perturbation theory describes weak physics remarkably well \cite{pdg}, even though it operates on the gauge-dependent elementary states of the Lagrangian, the Higgs, the $W$, the $Z$, and the fermion fields. It therefore appears as if the perturbative BRST quartet mechanism takes sufficient care of gauge-invariance \cite{Kugo:1979gm}. However, this is formally not correct. But the intricate structure of the standard-model and the previously mentioned FMS mechanism work together in such a way that corrections to the perturbative results are often negligible \cite{Frohlich:1980gj,Frohlich:1981yi,Maas:2017wzi}. This is not true in more general theories, and determining the consequences is the main aim of this work.

Before doing so, it is instructive to recapitulate why the BRST quartet mechanism is insufficient, and how this is ameliorated in the standard model. The origin of the problem is that the BRST quartet mechanism relies on BRST symmetry. However, beyond perturbation theory BRST symmetry is more subtle due to the Gribov-Singer ambiguity in a non-Abelian gauge theory \cite{Gribov:1977wm,Singer:1978dk,Fujikawa:1982ss,vanBaal:1991zw,DellAntonio:1991mms,vanBaal:1997gu,Vandersickel:2012tg,Maas:2011se}, and in general no longer realized in a way as to allow for the BRST quartet mechanism. In the case of Yang-Mills theory, the symmetry can likely be recovered at a non-perturbative level, at least in some subset of non-perturbatively extended gauges, but then implies the absence of all quantities carrying a gauge index from the physical spectrum, and thus especially all elementary particles \cite{Kugo:1979gm,Maas:2011se}. The same problem persists also in theories with a Brout-Englert-Higgs effect (BEH), but no similar resolution has been found so far \cite{Maas:2017wzi}. Even if such a resolution is possible, it will likely also imply the absence of all particles carrying a gauge charge from the physical spectrum. This is supported by various other, rather robust, field-theoretical statements implying the absence of all particles carrying a gauge charge from the spectrum. This includes results showing that the Gribov-Singer ambiguity obstructs a gauge-invariant dressing of elementary states \cite{Lavelle:1995ty}, the vanishing of any gauge-dependent expectation values \cite{Frohlich:1981yi}, and that the spectrum is qualitatively the same for both QCD-like and BEH-like physics in the standard-model Higgs sector \cite{Fischler:1974ue,Elitzur:1975im,Osterwalder:1977pc,Seiler:2015rwa,Banks:1979fi}. A review of these issues, including a list of frequently asked questions on this topic, can be found in \cite{Maas:2011se,Maas:2017wzi}.

The reason why it nonetheless seems to work in the standard model is subtle. It follows from a combination of the BEH effect and the FMS mechanism \cite{Frohlich:1980gj,Frohlich:1981yi}: Under certain conditions, met by the standard model, the properties of the physical states coincide with the ones determined using standard perturbation theory. This was confirmed in lattice simulations for the Higgs-weak sector \cite{Maas:2012tj,Maas:2013aia} and, in an exploratory manner, for the electroweak sector \cite{Shrock:1985un,Shrock:1985ur}. How it works will be sketched in Sec.~\ref{sec:recipie}. A brief review of this can be found in \cite{Torek:2016ede}, and in more detail in \cite{Maas:2017wzi}. While this also applies to the fermion sector in principle \cite{Frohlich:1980gj,Frohlich:1981yi,Egger:2017tkd,Maas:2017wzi}, the involved technical complications have yet prevented an explicit test. Still, the excellent agreement of the predictions of the FMS mechanism with experimental results is already good evidence for its validity.

The conditions mentioned are quite specific. Especially, the standard model is exceptional to fulfill them. The most important feature in this respect is that the local weak symmetry group is the same as the global custodial symmetry group of the Higgs. In general, BSM theories cannot be expected to accomplish this particular requirement \cite{Maas:2015gma}. If such a theory does not satisfy certain conditions, discrepancies between the actual physical spectrum and the one described by perturbation theory arise. Explicit investigations have shown both cases to occur \cite{Maas:2016qpu,Maas:2016ngo}. The most drastic consequence is that the (low-lying) observable spectrum is different than the one obtained in standard perturbation theory, and may no longer match the standard model. These would then be ruled out as new physics scenarios.

The aim of this work is to classify a set of theories according to their physical spectrum. This will be done using the FMS mechanism, as explained in Sec.~\ref{sec:recipie}. These results are analytical predictions, valid in a similar range as conventional perturbation theory. 
As long as no statistically reliable experimental indication of new physics arise, these predictions can be tested only theoretically, e.g., using lattice simulations along the lines of \cite{Maas:2012tj,Maas:2013aia,Maas:2016ngo}.

After introducing the general formalism we turn to explicit classes of theories. These are $\SU(N)$ gauge theories with a single Higgs in the fundamental representation in Sec.~\ref{sec:fundamental}, $\SU(N)$ gauge theories with a single Higgs in the adjoint representation in Sec.~\ref{sec:adjoint}, and finally GUT-like structures \cite{Bohm:2001yx} with one Higgs in the adjoint representation and one Higgs in the fundamental representation for $\SU(5)$ in Sec.~\ref{sec:gut}. This last section also shows that with increasingly complex structures in the Higgs sectors a full discussion quickly proliferates into an involved group-theoretical problem. This is especially true for multiple Higgs flavors in the same representation, as we show for the case of SU(3) with two Higgs in the fundamental representation in App.~\ref{sec:su32}, and as was already seen for the case of a 2HDM in \cite{Maas:2016qpu}.

\section{The general recipe}\label{sec:recipie}

In the following we will consider gauge theories coupled to scalar fields equipped with a potential that allows for a Brout-Englert-Higgs effect,
\begin{align}
 \mathcal{L} = -\frac{1}{2} \tr (F_{\mn}F^{\mn})  +  (D_{\m} \phi^{r}_{f})^{\dagger}_{\bar{a}} (D^{\m} \phi^{r}_{f})^{\phantom{\dagger}}_{\bar{a}}  -  V(\phi^{r}_{f}).
 \label{lagrangian}
\end{align}
The gauge fields $A_\mu$ with field-strength tensor $F_\mn$ couple through the covariant derivative $D_\m$ to the Higgs fields $\phi^{r}_{f}$, which are in some representation $r$ of the gauge group. 
For the Higgs potential $V$ we allow for any gauge-invariant scalar term built from the components of the scalar fields which is renormalizable by power-counting and which is required to have classically one or more minima at non-zero Higgs field.
We allow further for multiple Higgs fields in the same representation, counted by the index $f$. The precise structure of the potential will determine whether the theory contains global symmetries between the Higgs fields of a given representation, i.e., an enlarged global custodial symmetry. However, we will restrict the discussion to only one flavor in the fundamental representation or one flavor in the adjoint representation in the following sections, yielding $\U(1)$ and $\mathbb{Z}_2$ as custodial symmetry groups, respectively. We outline the strategy for multiple Higgs fields in the same representation in the appendix. 
The index $\bar{a}$ in the kinetic term has to be understood as a multi-index running over all possible components of the Higgs fields in a certain representation such that a gauge-invariant real scalar term is formed.

A physical observable can only be a gauge-invariant state\footnote{It should be noted that we use here state and operator partly interchangeable. This is done as we always only consider the state which is created by a given operator when acting on the vacuum, and thus the state and operator involved are uniquely related. Note, however, that these states are not necessarily only one-particle states, as will be seen already in expression (\ref{scalarsinglet}) below, where a single operator creates (at least) a superposition of a one-particle state, with a single Higgs, and a two-particle state with two Higgs particles. At the current level, this distinction is not crucial, as we can analytically map the operators to particular states and their particle content, since we do not consider interactions on the right-hand side and only operator bases with a single operator in every quantum number channel. Going beyond either restriction will require to make the distinction more carefully.}. In a non-Abelian gauge theory these are created by composite operators \cite{Haag:1992hx}, i.e., bound-state operators and not the elementary field operators present in the Lagrangian. These can be classified according to spin, (charge)parity, and custodial quantum numbers. 
The basic principle of the FMS mechanism \cite{Frohlich:1980gj,Frohlich:1981yi} is now straightforward.
The first step is to classify all possible gauge-invariant channels of interest with respect to these global quantum numbers.
A straightforward example is given by the operator $O_{0^{+}}(x) = (\phi_{\bar{a}}^{\dagger}\phi_{\bar{a}}^{\phantom{\dagger}})(x)$ which is the simplest operator in the  $0^{+}$ singlet channel.

The second step is to choose a gauge in which the Higgs fields acquire nonvanishing vacuum expectation values (vevs)  $\langle \phi \rangle = \phi_{0}$. This requirement is necessary to perform perturbative calculations within the FMS prescription. A convenient choice is given by the class of $R_{\xi}$ gauges which will be used throughout this work. We would like to emphasize at this point that whether the scalar field acquires a vev is a pure gauge choice and only within these gauges perturbation theory is applicable \cite{Lee:1974zg,Osterwalder:1977pc}.  Even if the Higgs potential has a Mexican-hat structure, gauges are possible in which the vev of the scalar field vanishes identically, e.g., by averaging over all possible minima of the potential \cite{Maas:2012ct}. Within such a gauge the mass of the elementary $W$ field would vanish identically to all orders in perturbation theory. However, the observable gauge-invariant vector state which characterizes the physical $W$ remains massive due to nonperturbative effects.

All physical relevant information of an operator is stored in its $n$-point function. In order to extract the mass, we can investigate the propagator $\langle O^{\dagger}(x)O(y)\rangle$. 
The last step of the FMS prescription is to expand the Higgs fields in fluctuations $\varphi$ around the gauge-dependent vev, $\phi(x) = \phi_{0} + \varphi(x)$, within the gauge-invariant correlators. This yields 
\begin{align}
 \big\langle O_{0^{+}}^\dagger(x) O_{0^{+}}(y)\big\rangle  &=  \big\langle \mathrm{Re}\big( \phi_{0\, \bar{a}}^\dagger\, \varphi_{\bar{a}}^{\phantom{\dagger}} \big)^{\dagger}(x) \;\mathrm{Re}\big( \phi_{0\, \bar{a}}^\dagger\, \varphi_{\bar{a}}^{\phantom{\dagger}} \big)(y) \big\rangle \notag \\ 
 &\quad + \cdots,
 \label{scalarsinglet}
\end{align}
for our simple example in the $0^{+}$ singlet channel.
The right-hand side is just an ordinary fluctuation-field propagator corresponding to the propagator of an elementary Higgs excitation in a given representation and flavor along the radial direction of the vev. 
Comparing poles on both sides implies that the physical scalar has the same mass as the elementary fluctuation field. 
This explains why the observable particle has the same mass as the elementary Higgs field in the standard model, and is thus well described by perturbation theory. This consideration can be also extended to the full standard model \cite{Frohlich:1980gj,Frohlich:1981yi} and has been confirmed in explicit lattice calculations for some example theories \cite{Maas:2012tj,Maas:2013aia,Maas:unpublishedtoerek}. 

The neglected parts in Eq.~\eqref{scalarsinglet} can also contribute to the spectrum of the $0^{+}$ operator. Whether these additional contributions are further bound states, resonances or scattering states depends on the actual considered theory.
Further, Eq.~\eqref{scalarsinglet} stresses the necessity to choose a gauge with a nonvanishing vev for the scalar field in order to use the FMS prescription to predict the mass. For a vanishing vev, the FMS expansion would trivially reproduce the original bound-state operator which leads, of course, to a true but non-illuminating statement.

Because of gauge invariance there are also no gauge multiplets in the physical spectrum. The only possible multiplets arise if there are global symmetries. Then, the physical states can be multiplets in such global symmetries. In this work, we consider only custodial multiplets. If other global symmetries are present, this may also involve these symmetries \cite{Maas:2015gma}.

The particular consequences of a custodial symmetry depend on the specific example. For instance, in the standard model, with its SU(2) custodial symmetry, a custodial triplet with spin 1 is found to have the same mass as the $W$ and $Z$ bosons, forming their physical equivalents \cite{Frohlich:1980gj,Frohlich:1981yi}, which has also been confirmed on the lattice \cite{Maas:2012tj,Maas:2013aia}. As will be seen in the following, the global multiplet structure plays a central role for the particle spectrum, as has already been seen in explicit examples \cite{Maas:2015gma,Maas:2016ngo,Maas:2016qpu}. It also plays an important role in the fermion sector \cite{Frohlich:1980gj,Frohlich:1981yi,Torek:2016ede,Egger:2017tkd}, though this sector will not be considered here.

While the FMS expansion is always possible, it does not necessarily lead for every operator and/or every quantum number channel to a single elementary particle correlator at leading order as in Eq.~\eqref{scalarsinglet}. 
In fact, the first non-vanishing term can be a scattering state for some operators, i.e., a state involving two or more elementary fields. Such operators may yield, just as in the quark model, an additional bound state mass made up from the combined masses of the particles in the scattering state. We will come back to this in the following sections.

In principle, the FMS mechanism can be expected to work well if the gauge-dependent propagators are described well by perturbation theory for any gauge with a non-vanishing Higgs condensate. This condition can, unfortunately, not be self-consistently guaranteed in perturbation theory alone, as has been seen in explicit lattice calculations \cite{Maas:2014pba}. However, this is exactly the same kind of condition applying to ordinary perturbation theory.

In the remainder of this work we use this general recipe to determine the physical spectrum of different classes of theories, as well as for some particular explicit examples.

Before doing so, it is worthwhile to point out two particulars with respect to the class of operators considered to describe the physical particles. These are composite, local operators. Each of the two qualifiers deserves a few words.

First, we consider only operators local in the sense that they depend only on a single space-time argument. Thus so do the created states. This is motivated by the fact that we want to describe objects characterized by a space-time position. Still, the operators themselves can contain parallel transporters, sensitive to the local surroundings of the space-time point, e.\ g.\ (\ref{eq:fund1-}), or even involving non-local structures like (\ref{eq:vectorZ2even}) and (\ref{eq:vectorZ2odd}). At the lowest-order perturbative expansion the states created by the operators we consider are indeed localizable particles. Beyond that, the states become harder to characterize, as they may describe extended or more complicated objects \cite{Lavelle:1995ty,Haag:1992hx,Osterwalder:1977pc}. We thus assume that considering operators of the described classes do indeed describe localizable particles and are sufficient to describe all particle-like excitations in any given quantum number channel. Concerning the status for other operator classes in theories of the type considered here, we refer to \cite{Greensite:2017ajx,Greensite:2018mhh} for recent investigations.

Second, the operators are composite of field operators at the same space-time point. This implies additional renormalization \cite{Itzykson:1980rh} compared to ordinary field operators. However, we expect that this ultimately works like for operators which, e.g., are used to describe hadrons in QCD. Investigations of such operators numerically using lattice regularization supports this assumption \cite{Shrock:1985un,Shrock:1985ur,Lee:1985yi,Maas:2012tj,Maas:2013aia,Maas:2016ngo,Maas:unpublishedtoerek}. As we treat the right-hand side correlation functions perturbatively, standard methods for composite operators are sufficient \cite{Itzykson:1980rh}. For instance, an explicit calculation of (\ref{scalarsinglet}) shows that disconnected contributions need to be explicitly regularized and subtracted. Fortunately, this does not affect the pole structure, leaving the spectroscopical results which are the main focus here, untouched. An exact non-perturbative determination of the gauge-dependent composite correlation functions on the right-hand side of the expansion has, to our knowledge, so far not been performed. But as these describe effectively gauge-dependent bound-states, we suspect that the same applies to them as for the left-hand side.


\section{SU(N) gauge theories with a fundamental Higgs}
\label{sec:fundamental}

We consider an SU($N>2$) gauge theory coupled to a single scalar field in the fundamental representation.\footnote{See \cite{Frohlich:1980gj,Frohlich:1981yi,Maas:2013aia,Maas:2016qpu} for the SU(2) fundamental case.} This is a special case of the Lagrangian~\eqref{lagrangian}, where the Higgs field $\phi$ denotes a complex $N$-component vector transforming as $\phi(x) \to U(x) \phi(x)$ with $U(x) = \E^{\I T_{i}\alpha_{i}(x)} \in \SUN$  and $T_{i}$ are the generators of the associated Lie algebra. The latter can be constructed explicitly by the generalized Gell-Man matrices for the fundamental representation. The Higgs potential depends only on the invariant $\phi^{\dagger}\phi$ and the scalar kinetic term reads $(D_{\mu}\phi)^{\dagger}(D^{\mu}\phi)$.

\subsection{Gauge-variant description in a fixed gauge}

The precise shape of the scalar potential determines whether a non-zero (unique) vacuum expectation value is classically possible.\footnote{We assume here and in the following always implicitly that this is still possible at the quantum level. This is not guaranteed, as lattice calculations have shown explicitly \cite{Maas:2014pba}. However, only in this case the gauge condition (\ref{gaugec}) can be chosen and the FMS mechanism meaningfully applied, as discussed in Sec.~\ref{sec:recipie}.} We implement the convenient $R_{\xi}$ gauge condition\footnote{The issue of the Gribov-Singer ambiguity \cite{Gribov:1977wm,Singer:1978dk} is likely quantitatively irrelevant for models with sufficiently large Higgs condensates in case all non-Abelian gauge bosons acquire a mass term, see \cite{Lenz:2000zt,Capri:2013oja} as well as \cite{Maas:2010nc} for an explicit lattice calculation. Nonetheless, we expect some remnant of the Gribov-Singer ambiguity as a non-Abelian subgroup remains unbroken in these type of models. In addition, the qualitative effects, like the breaking of perturbative BRST allured to in the introduction, remain.} 
\begin{align}
 \mathcal{L}_{\mathrm{gf}}  =  -\frac{1}{2\xi} \Big|  \pdm A^{\mu}_{i} + \frac{\I g v \xi}{\sqrt{2}} (n^{\dagger} T_{i} \varphi - \varphi^{\dagger} T_{i} n) \Big|^{2},
 \label{gaugec}
\end{align}
with $n$ a unit vector, $n^{\dagger}n=1$, in the fundamental representation of the gauge group. Its direction defines the direction of the vev, and $v$ is its absolute value, $\phi_{0}^{\dagger}\phi_{0}^{\phantom{\dagger}} = v^{2}/2$. 
The vacuum expectation value satisfies $\pd_{\phi} V|_{\phi =\phi_{0}=\frac{v}{\sqrt{2}}n } = 0$. 
In addition, in $R_{\xi}$ gauges the would-be Goldstone bosons also obtain a mass at tree-level. 
This mass is proportional to the gauge fixing parameter, but any effects will be removed due to cancellations between the would-be Goldstone bosons, the time-like gauge bosons, and the ghosts. Thus, they drop out in any vacuum correlator \cite{Bohm:2001yx}. They will therefore play no role in the following.

To investigate the mass spectrum at tree-level in this setting the vev of the scalar is split off the fluctuation part $\varphi$,
\begin{align}
 \phi (x) 
 = \frac{v}{\sqrt{2}} n + \varphi(x).
 \label{eq:SplitHiggsF}
\end{align}
The spectrum contains one real-valued massive scalar degree of freedom and $2N-1$ would-be Goldstone modes. The non-Goldstone Higgs boson which is the excitation of the scalar field along $n$, as well as the would-be Goldstones can be described in a gauge-covariant (but not gauge-invariant) manner without specifying $n$ by $h \equiv \sqrt{2}\, \re(n^{\dagger}\phi)$ and $\breve{\varphi} \equiv \phi - \re(n^{\dagger}\phi)n = \varphi - \re(n^{\dagger}\varphi)n$, respectively. In the following the explicit choice $n^{a} = \delta^{aN}$ is usually made. Without loss of generality this is always possible by a local gauge transformation. Thus, the vev is always in the real part of the $N$th component.\footnote{This might change once nonperturbative solutions of the equations of motion of the elementary fields like instantons are considered as well and/or in the presence of gauge defects. For simplicity, we do not consider these type of additional nonperturbative effects in the following.} 
Note that we will use the following convention from now on. Upper indices starting from $a,b,c,\cdots$ will indicate components from an object in the fundamental representation of the gauge group while lower indices starting from $i,j,k,\cdots$ will denote adjoint indices.

Rewriting the scalar kinetic term in the Lagrangian \eqref{lagrangian} by splitting the Higgs field into the vev and the fluctuation part, Eq.~\eqref{eq:SplitHiggsF}, we obtain
\begin{align}
\begin{split}
 (D_{\m} \phi)^{\dagger} (D^{\m} \phi) &= \pdm \varphi^{\dagger} \pd^{\m} \varphi + \frac{g^{2}v^{2}}{2}\, n^{\dagger} T_{i}T_{j} n \,  A^{\mu}_{i}A_{j\, \mu} \\ 
 &\quad + \sqrt{2}gv\, \im(n^{\dagger} T_{i}\pd_{\mu} \varphi)A_{i}^{\mu}  + \cdots\;.
\end{split}
\label{eq:CovDev}
\end{align}
This includes the usual \cite{Bohm:2001yx} mass matrix for the gauge bosons in the first line. In the second line the mixing between the longitudinal parts of the gauge bosons and the Goldstone bosons appear. The neglected part describes the three and four point vertices between the scalar and gauge field. 
Only the massive gauge bosons mix with the Goldstone bosons. By choosing a suitable gauge condition these mixing terms are removed, which we will always do. Thereby the additional part proportional to the Goldstone fields is exactly constructed in such a way as to obtain a propagator diagonal in field space.

The mass matrix $(M^{2}_{A})_{{ij}}$ of the gauge bosons is already diagonal for our convenient choice of the direction of the vev, $n^{a}=\delta^{aN}$, and is given by, 
\begin{align}
\begin{split}
 (\MA^{2})_{ij} &= \frac{g^{2}v^{2}}{2}\, n^{\dagger} \{T_{i},T_{j}\} n  \\
 &=  \frac{g^{2}v^{2}}{4} \diag\Big( \underbrace{0,\cdots,0}_{(N-1)^{2}-1} , \underbrace{1,\cdots,1}_{2N-2},\frac{2}{N}(N-1)\Big)_{ij}.
 \end{split}
\end{align}
Thus, there are $(N-1)^{2}-1$ massless gauge bosons, $2N-2$ degenerated massive gauge bosons with mass $\mA = \frac{1}{2}gv$ and one with mass $\MA = \sqrt{2(N-1)/N}~\mA$. 
Moreover, the elementary Higgs field has a mass $\mh^{2} = \lambda^{2}v^{2}$, with $\lambda$ being the four-Higgs coupling, $\frac{\lambda}{2} (\phi^{\dagger}\phi)^{2}$.

In an abuse of language, this establishes the situation which is commonly called 'spontaneous breaking' in case the system is in the Brout-Englert-Higgs phase.
The breaking pattern reads $\SU(N) \to \SU(N-1)$. 
With respect to the subgroup SU($N-1$) the gauge bosons are in the adjoint representation (massless), a fundamental and an anti-fundamental representation (mass $\mA$) and a singlet representation (mass $\MA$), explaining their degeneracy pattern.

\subsection{Gauge-invariant spectrum}
\label{fundgi}

So far, the construction was the standard perturbative one. We now turn to the observable gauge-invariant states of the theory and apply the FMS mechanism to predict the physical mass spectrum.
We construct the gauge-invariant spectrum according to the multiplets of the global symmetries of the theory. The Lagrangian exhibits a global $\mathrm{U}(1)$ symmetry, acting only on the scalar field, besides the local $\SUN$ gauge symmetry.

It is straightforward to construct a gauge-invariant state for the Higgs boson, which should be a scalar with positive parity and a singlet under the global symmetry, $J^{P}_{\U(1)} = 0^{+}_{0}$. Therefore, we consider the following gauge-invariant composite operator,
\begin{align}
 O_{0^{+}_0}(x) = \big( \phi^{\dagger}\phi \big)(x),
 \label{eq:giHiggs}
\end{align}
which exhibits the demanded quantum numbers. We apply the FMS prescription according to Sec.~\ref{sec:recipie}, to predict the mass of the state provided by the mapping from the gauge-invariant to the elementary operators. The expanded correlation function in the scalar fluctuations reads in detail
\begin{align}
 \big\langle O^\dagger_{0^{+}_0}(x)O_{0^{+}_0}(y)\big\rangle  &=      
 \frac{v^{4}}{4}  +\frac{v^{3}}{2} \big\langle h(x)  +  h(y) \big\rangle \notag \\ 
 &\quad + v^{2} \big\langle h(x)h(y) \big\rangle  \notag \\
 &\quad + \frac{v^{2}}{2} \big\langle \big(\varphi^{\dagger}\varphi\big)(x)  +  \big(\varphi^{\dagger}\varphi\big)(y) \big\rangle \notag \\
 &\quad+ v \big\langle h(x) \big(\varphi^{\dagger}\varphi\big)(y) + \big(\varphi^{\dagger}\varphi\big)(x)h(y)\big \rangle \notag  \\
 &\quad+ \big\langle \big(\varphi^{\dagger}\varphi\big)(x) \big(\varphi^{\dagger}\varphi\big)(y)\big \rangle.
 \label{eq:fund0+}
\end{align}
Although every individual term on the right-hand side is a gauge-variant object, the sum of all terms is gauge invariant. So far, no approximation has been applied and Eq.~\eqref{eq:fund0+} is an exact identity. The remaining task is to calculate the properties of the correlation functions. While the left-hand side involves the propagation of a gauge-invariant but complicated bound state, the FMS mechanism provides a mapping to gauge-dependent elementary correlation functions in a fixed gauge. Approximations for the latter can be calculated by a variety  of tools, e.g., perturbation theory, lattice, or functional methods. 

For the moment, we will restrict the calculations to the simplest possible approximation, a tree-level analysis.
Computing the $n$-point functions on the right-hand side at tree-level (tl), we get
\begin{align}
\begin{split}
 \big\langle O^\dagger_{0^{+}_0}(x) O_{0^{+}_0}(y)\big\rangle &=   v^{2} \big\langle h(x)h(y) \big\rangle_{\mathrm{tl}}  
 +\big\langle h(x)h(y)\big \rangle^{2}_{\mathrm{tl}} \\
 &\quad + \frac{v^{4}}{4} +\mathcal{O}(\varphi^{3},g,\lambda),
\end{split}
\end{align}
where the first term describes the propagation of a single elementary Higgs boson from $x$ to $y$. 
The second term describes two propagating non-interacting elementary Higgs bosons both starting at $x$ and ending at $y$, which is a scattering state. This implies that the right-hand side has one pole, at the tree-level Higgs mass, and a cut starting at twice the tree-level Higgs mass. 
Comparing poles on both sides, this procedure predicts that the physical, gauge-invariant left-hand side should have a mass equal to the tree-level mass of the elementary Higgs, $m_{0^{+}_0}=\mh=\sqrt{\lambda}v$, and the next state should then be the trivial scattering state of twice this mass. For the SU(2) case and the SU(3) case both features have been confirmed on the lattice \cite{Maas:2012tj,Maas:2013aia,Maas:2014pba,Maas:unpublishedtoerek}. Moreover, additional bound states with mass $\mh < m < 2\mh$ may exist beyond the simple tree-level analysis, depending on the analytic structure of the full $2$-point function $\langle h(x)h(y)\rangle$ on the right hand side of Eq.~\eqref{eq:fund0+}.\footnote{However, at least for the SU(2) case, where the same analysis applies, no signal of such states has been seen in lattice calculations \cite{Maas:2014pba,Wurtz:2013ova}. But the predicted scattering states have been seen, necessarily.}

Next we construct a singlet vector operator and apply the FMS procedure:
\begin{align}
 O^{\mu}_{1^{-}_0}(x)  =   \I\, \big( \phi^{\dagger} D^{\mu} \phi \big) (x)  =   -\frac{v^{2}g}{2}  \big( n^{\dagger}A^{\mu}n \big)(x)  + \mathcal{O}(\varphi).
 \label{eq:fund1-}
\end{align}
The correlator is given by
\begin{align}
 & \big\langle O^{\mu\,\dagger}_{1^{-}_0}(x) O^{\nu}_{1^{-}_0}(y) \big\rangle  \notag \\
 &\stackrel{\hphantom{\;\; n^{a}=\delta^{Na}}}{=}  \frac{v^{4}g^{2}}{4} \big\langle n^{\dagger} A^{\mu}(x) nn^{\dagger} A^{\nu}(y) n \big \rangle + \mathcal{O}(\varphi)  \\
 &\stackrel{\;\; n^{a}=\delta^{Na}}{=} \frac{(N-1)v^{4}g^{2}}{8N} \big\langle A_{N^2-1}^{\mu}(x) A_{N^2-1}^{\nu}(y)\big \rangle + \mathcal{O}(\varphi) \notag
\end{align}
to leading order in the FMS expansion.
We project generically on the heaviest elementary state due to the projector $nn^{\dagger}$ between the gauge fields in the propagator. Thus, this predicts a single vector particle with the mass of the heaviest gauge boson, $m_{1^{-}_0}=\MA$. For the case of $\SU(3)$, this has been confirmed in lattice calculations \cite{Maas:2016ngo,Maas:unpublishedtoerek}.\footnote{Note that the situation for SU(2) is different because of the differing global symmetry, but also there the results from the FMS mechanism have been confirmed on the lattice \cite{Maas:2012tj,Maas:2013aia}.} Here, we predict this to be true for any $N \geq 3$. 

Besides the gauge-invariant prescription of the observable Higgs boson with the operator \eqref{eq:giHiggs} which has the mass of an elementary Higgs field, we also obtain a state which can be mapped on an elementary gauge boson in the $1^{-}$ channel. Our analysis for the simplest operator in the $0^{+}$ channel might tempt to the conclusion that it is sufficient for the calculation of the ground state mass to truncate the FMS expansion after the first nontrivial elementary operator which leads to an elementary propagator on the right-hand side. For the $0^{+}$ channel this is the elementary Higgs field $O_{0^{+}_0}(x) = v^{2}/2 + v h(x) + \mathcal{O}(\varphi^{2})$ and propagators coming from the neglected parts indeed contribute only to scattering states. 
However, the situation can be more involved for more intricate bound states in other channels.

%
In order to illustrate this, we take the next-to-leading order of the FMS prescription for the vector operator \eqref{eq:fund1-} into account. At this level, the vector operator reads 
\begin{align}
 O^{\mu}_{1^{-}_0}  = &-\frac{v^{2}g}{2}  \big( n^{\dagger}A_{\mu}n \big) + \I \frac{v}{\sqrt{2}}\, n^{\dagger}\pdm \varphi \notag \\ 
 &-\sqrt{2}g v \mathrm{Re}(n^{\dagger} A^{\mu}\varphi) + \mathcal{O}(\varphi^{2}).
\end{align}
While the terms separated in the second line indeed give scattering states of the elementary Higgs with the massive gauge bosons once the correlator of the $1^{-}_{0}$ operator is investigated, the first term linear in $\varphi$ also contributes to the pole structure of the gauge-invariant operator. 
Thus, the actual propagator of the vector operator $O_{1^{-}_{0}}$ reads to lowest order in the elementary $n$-point functions:
\begin{align}
 \big\langle O^{\mu\, \dagger}_{1^{-}_0}(x) O^{\nu}_{1^{-}_0}(y) \big\rangle 
 &= \frac{(N-1)v^{4}g^{2}}{8N} \big\langle A_{N^2-1}^{\mu}(x) A_{N^2-1}^{\nu}(y)\big \rangle_{\mathrm{tl}} \notag\\ 
 &\quad + \frac{v^{2}}{2}\partial_{x}^{\mu}\partial_{y}^{\nu} \big\langle h(x) h(y) \big\rangle_{\mathrm{tl}}  +  \cdots,
\end{align}
where the $\cdots$ contain the $(n\geq 3)$-point functions. 
Identifying poles on the right-hand side, we get a pole at $\MA$, the mass of the heaviest gauge boson. But also one at the mass of the elementary Higgs field $\mh$ arises to this order in the FMS expansion.

Nonetheless, this additional pole structure does not necessarily imply that an additional particle in the vector channel is predicted as it does not exhibit the expected Lorentz structure of a massive vector boson as this pole appears only in the longitudinal part of the correlator. It rather reflects the fact that a derivative acting on a scalar operator transforms as a vector and therefore mixes with vector operators in the $1^{-}$ channel but does not indicate a next-level state in this channel, at least to this order in the approximation. In general a detailed variational analysis including various operators with a sufficiently large overlap with the ground as well as the first few excited states within a specific channel is required to make a definite statement about potential higher excitations.




Nevertheless, higher-order terms in the fluctuating elementary fields can in principle also lead to nonvanishing $2$-point functions which contribute to the spectrum of an operator in certain cases even if this was not the case in the previous example. 
Whether this leads to an additional particle either in terms of an additional bound state or a resonance, or just a nontrivial scattering state within the considered channel, depends on the precise properties of the model.\footnote{For the rest of this paper, we will use the term \textit{next-level state} in case the state is not the ground state or a trivial scattering state of $n$ times the mass of the ground state. As more sophisticated analyses beyond these simple considerations are needed to identify whether this state is an additional bound state, a scattering state, or a resonance, we will simply stick to the term next-level state, keeping in mind that this state can also predict an additional particle in case it is a bound state or a resonance.}
We illustrate such an example in the following.


\begin{table*}[t!]
\begin{center}
\begin{tabular}{c|ccc|ccccc}

\multicolumn{4}{c}{elementary spectrum} & \multicolumn{4}{c}{gauge-invariant spectrum} \cr

\toprule

$J^P$ & \quad Field \quad & \quad Mass \quad & \quad Degeneracy \quad & \quad $\U(1)$ \quad & \quad Operator \quad & \quad Mass \quad & \quad Next-level state \quad & \quad Degeneracy \quad \cr
\hline

$0^+$ & $h$ & $\mh$ & $1$ & $0$ & $O_{0^+_0}$ & $\mh$ & - & $1$\cr
$\quad$ & $\quad$ & $\quad$ & $\quad$ & $\pm 1$ & $O_{0^+_{\pm 1}}$ & \quad $(N-1)\mA$ \quad & \quad $(N-1)\mA + \MA$ \quad & $1/\bar{1}$\cr
\hline
$1^-$ & $A^\mu_{1,\dots,(N-1)^2-1}$ & $0$ & $(N-1)^2-1$ & $0$ & $O^\mu_{1^-_{0}}$ & $\MA$ & - & $1$\cr
 & $A^\mu_{(N-1)^2,\dots,N^2-2}$ & $\mA$ & $2(N-1)$ & $\pm 1$ & $O^\mu_{1^-_{\pm 1}}$ & $(N-1)\mA$ & \quad $(N-1)\mA + \MA$ \quad & $1/\bar{1}$ \cr
 & $A^\mu_{N^2-1}$ & $\MA$ & $1$  &  &  &  & & \cr
\toprule
\end{tabular}
\caption{Left: Gauge-variant spectrum of an $\SU(N)$ gauge theory with a single scalar field in the fundamental representation. Right: Gauge-invariant (physical) spectrum of the theory. Here $\mh$ denotes the mass of the elementary Higgs field, $\MA$ is the mass of the heaviest elementary gauge boson and $\mA$ the mass of the degenerated lighter massive gauge bosons. 
We assign a custodial $\U(1)$ charge of $1/N$ to the scalar field $\phi$.
The column \textit{next-level state} lists the masses of possible additional bound states or resonances, see the discussion in the main text and in App.~\ref{sec:full}.  Whether these states are indeed bound states or resonances or only nontrivial scattering states can only be decided once the full analytical structure of the correlator is investigated. Trivial scattering states are ignored.
The definition of the fields and operators can be found in the main text.}
\label{tab:suf}
\end{center}
\end{table*}


Besides the $\mathrm{U}(1)$ singlet states, we can also construct states with open $\mathrm{U}(1)$ quantum numbers. This is important as the lightest such state is absolutely stable in the theory, as this charge is conserved. 

For $\SU(3)$ an explicit example for a vector state is given by
\begin{align}
 O^\mu_{1^-_1} (x) = \Big[\epsilon^{abc} \phi^{a} (D^{\mu}\phi)^{b} (D^{2}\phi)^{c}\Big](x),
 \label{eq:Op-open}
\end{align}
where we assigned a $\U(1)$ charge $1/N = 1/3$ to the scalar field $\phi$.
To leading order in the FMS mechanism as well as in perturbation theory, the correlator of this operator expands to a product of three propagators. 

As a simple-minded constituent model, we interpret the mass of the gauge-invariant operator by the sum of the masses of the three propagators on the right-hand side of the expansion.

The mass of this state is given by $m_{1^-_1} = 2\mA$: In leading order the composite state is described by three independently propagating gauge bosons. One from the massless elementary gauge boson and two gauge bosons with mass $\mA$, giving the total mass. Likewise, a second, non-interacting state is admixed with mass $m_{1^-_1}^{*} = 2\mA + \MA$. Of course, in addition there are corresponding anti-particles of the same mass but opposite (U(1)) charge described by $O^\mu_{1^-_{-1}} = {O^\mu_{1^-_1}}^{\dagger}$. Lattice investigations of the masses of these states for $N=3$ support \cite{Maas:unpublishedtoerek} this non-trivial prediction.

Besides the presented gauge-invariant operators, it is of course also possible to construct other operators within a specific channel. We expect that those operators have the largest overlap with the ground state of a given channel, which have the least field content. For the $\U(1)$-singlet operators the unique solution to this requirement is given. However, there are ambiguities regarding the other channels. For instance, the operators $\epsilon^{abc} \phi^{a} (D^{\nu}\phi)^{b} (D_{\{\nu}D_{\mu\}}\phi)^{c}$ and $\epsilon^{abc} \phi^{a} (D^{\nu}\phi)^{b} (F_{\nu\mu}\phi)^{c}$ equally describe a vector state with nonvanishing $\U(1)$ quantum number. The advantage of the latter is that it can be straightforwardly generalized to an arbitrary $\SU(N)$ theory.
An explicit prescription for these operators is given by
\begin{align}
  O^\mu_{1^-_1}  =  \epsilon^{a_{1}\cdots a_{N}} \phi^{a_{1}}(D_{\nu_{1}}\phi)^{a_2} (F^{\nu_{1}}_{\phantom{\nu_1}\nu_{2}}\phi)^{a_{3}} \cdots (F^{\nu_{N-2}\mu} \, \phi)^{a_N}.
  \label{eq:fund-U(1)vec}
\end{align}
It is straightforward to convince oneself that the lightest pole of this operator is given by $(N-1)\mA$ for any $N>2$. 
Moreover, we get several excited states. For instance a next-level state is predicted with mass $(N-1)\mA + \MA$ as well as a trivial scattering state with mass $(N-1)\mA + \mh$.
A sketch of the analysis of the FMS description of these operators can be found in App.~\ref{sec:full}.

In contrast to the scalar channel for the elementary fields, we can also construct scalar operators with an open $\U(1)$ quantum number. These read for instance
\begin{align}
\begin{split}
  O_{0^+_1}  &=  \epsilon^{a_{1}\cdots a_{N}}  \phi^{a_{1}} (D_{\mu_{1}}\phi)^{a_2} (F^{\mu_{1}}_{\phantom{\mu_1}\mu_{2}}\phi)^{a_{3}} \cdots \\
 &\qquad\qquad\quad \cdots (F^{\mu_{N-3}}_{\phantom{\nu_{N-3}}\mu_{N-2}} \,\phi)^{a_{N-1}} (D^{\mu_{N-2}}\phi)^{a_N},
\end{split}
\label{eq:fund-U(1)sca}
\end{align}
for $N>3$ and $\epsilon^{a_{1}a_{2}a_{3}} \phi^{a_{1}} (D_{\mu}\phi)^{a_2} (D_{\nu}F^{\mu\nu}\phi)^{a_{3}}$ for ${N=3}$ and we predict them to have a ground-state mass of $m_{0^+_1} = (N-1)\mA$ as well as a mass of $(N-1)\mA + \MA$ for the next-level state for $\MA<\mh$.

Summarizing this section, we observed a qualitative difference of the elementary spectrum and the gauge-invariant observable spectrum for $\SU(N>2)$ for the class of theories described here. The difference are explicitly visible from the summary Tab.~\ref{tab:suf}. The elementary spectrum, which coincides with the one predicted by ordinary perturbation theory, contains one particle with mass $\mh$ in the $0^{+}$ channel, while in the gauge-invariant case there are least three scalar states. One of them is a singlet under the global $\U(1)$ symmetry group, and has mass $\mh$. The other two are a massive particle-anti-particle pair with non-vanishing $\U(1)$ quantum number. Also the vector channel differs. The elementary spectrum has $(N-1)^{2}-1$ massless gauge bosons, degenerated $2(N-1)$ massive ones and one gauge boson with a generically heavier mass. The gauge-invariant vector states are given by a single $\U(1)$-singlet and a massive particle-antiparticle pair with open $\U(1)$ quantum number. Moreover, it is possible that more massive particles in the gauge-invariant spectrum arise which can manifest as further bound states or resonances. We would like to emphasize, that the degrees of freedom of the gauge-invariant spectrum do not change once the gauge group is altered. Only the masses and other physical properties like decay constants change, in contrast to the elementary spectrum where also the number of degrees of freedom increases once $N$ increases. Explicit lattice simulations for the $N=3$ case support the results presented in this section \cite{Maas:2016ngo,Maas:unpublishedtoerek}.

\section{SU(N) gauge theories with an adjoint Higgs}
\label{sec:adjoint}

After the detailed discussion of the fundamental case, we now consider an $\SUN$ gauge theory coupled to a scalar field in the adjoint representation. In this case, it is useful to formulate the Lagrangian (\ref{lagrangian}) as
\begin{align}
 \mathcal{L} = -\frac{1}{4} F_{i\,\mn}F^{\mn}_{i}  +  \tr [(D_{\m} \Sigma)^{\dagger} (D^{\m} \Sigma)]  -  V(\Sigma).
 \label{eq:LagHiggsA}
\end{align}
The scalar field $\Sigma = \Sigma_{i}T_{i}$ transforms as $\Sigma(x) \to U(x) \Sigma(x) U(x)^{\dagger}$ and its components $\Sigma_{i}$ form a $N^{2}-1$ dimensional real-valued vector. The covariant derivative acting on $\Sigma$ is given by $D^{\m} \Sigma = \pd^{\mu} \Sigma + \I g [A^{\mu},\Sigma]$. The scalar potential $V$ contains all possible couplings up to fourth order in the scalar field spanned by the invariant Casimirs of the gauge group
\begin{align}
 V = -\mu^{2}\, \tr\Sigma^{2} + \gamma\, \tr \Sigma^{3} + \frac{\lambda}{2}(\tr\Sigma^{2})^{2} + \tilde{\lambda}\, \tr\Sigma^{4}.
 \label{eq:potadj}
\end{align}
For most of the time, we will enforce the action to be invariant under a discrete $\mathbb{Z}_{2}$ symmetry, i.e., $\gamma=0$. The case of a nonvanishing $\gamma$ will be exemplified in Sec.~\ref{sec:SU(3)adjoint}. 

It is important to note that the adjoint case induces a very different structure than the fundamental case where the only little group is $\SU(N-1)$ and all minima of the potential belong to the same gauge orbit \cite{O'Raifeartaigh:1986vq}. By contrast, different directions of the vev can belong to different little groups and thus to different physical theories with different mass spectra for the elementary fields as not all directions of the vev can be connected via a gauge transformation for a Higgs in the adjoint representation. This also induces further subtleties in the FMS prescription for the gauge-invariant spectrum which is discussed in more detail in App.~\ref{a:state}.

\subsection{Gauge-variant description in a fixed gauge}
\label{sec:adjoint-gv}

Assuming that the (effective) potential allows for a Brout-Englert-Higgs effect and choosing a gauge with a nonvanishing vev, we again split the scalar field in its vev and fluctuations around it:
\begin{align}
 \Sigma(x) = \langle\Sigma\rangle  +  \sigma(x) \equiv w\, \Sigma^{0}  +  \sigma(x).
\end{align}
The breaking pattern depends on the direction of the vev $\Sigma^{0}_{i}$ with $\Sigma^{0}_{i}\Sigma^{0}_{i}=1$. We again implement the gauge condition such that it removes the mixing between the gauge bosons and the would-be Goldstones, leading to a gauge-fixing Lagrangian
\begin{align}
 \mathcal{L}_{\mathrm{gf}} = \frac{1}{\xi} \tr\, (\pdm A^{\mu} + \I g \xi [\Sigma^{0},\Sigma])^{2}.
 \label{eq:gaugeAdj}
\end{align}
We can always choose a gauge in which  $\Sigma^{0}$ is diagonal due to the fact that every unitary matrix can be diagonalized by a suitable unitary transformation. Thus, it is sufficient to consider vevs spanned by the generators of the Cartan subalgebra as every element of the $\SU(N)$ algebra can be rotated to an element of the diagonal generators. 

The mass matrix for the gauge fields is given by
\begin{align}
 (\MA^{2})_{ij}  =  -2(gw)^{2}\, \tr\big( [T_{i}, \Sigma^{0}] [T_{j}, \Sigma^{0}]\big).
\end{align}
Whether a gauge boson acquires a mass or remains massless depends on whether the generator which is associated to that gauge boson commutes with $\Sigma^{0}$.

To identify all possible breaking patterns for a given group $\SU(N)$ corresponds to the identification of all possible little groups. This can be mapped on the combinatorial problem of finding all partitions $p(N)$ of the number $N$. The function $p(N)$ can be extracted from the formal Taylor series of the inverse Euler function $\prod_{k=1}^{\infty} 1/(1-x)^{k} = \sum_{N=0}^{\infty}p(N)x^{N}$. However, when it comes to the actual minimization of the potential energy of the scalar field, it can be shown that the potential has extrema only if  the vev $\Sigma^{0}$ has at most two different eigenvalues \cite{Li:1973mq,Ruegg:1980gf,Murphy:1983rf}. Therefore, the only relevant breaking patterns which can lead to a minimum of the potential are $\SU(N) \to \mathrm{S}\big( \U(P)\times \U(N-P) \big)$ with $P<N$ and one only has to consider $\lfloor N/2 \rfloor$ breaking patterns.\footnote{Moreover, it can be shown that the global minimum of the potential is given by the breaking pattern $P=\lfloor N/2 \rfloor$ for $\tilde{\lambda}>0$ and $P=1$ for $\tilde{\lambda}<0$ \cite{Li:1973mq}. Nevertheless, we will determine the spectrum also for the minima in which the theory is in a metastable state as this minima can become global once other fields are coupled to the theory.}

Thus, we obtain $N^{2}-1-2P(N-P)$ massless and $2P(N-P)$ massive gauge fields in the spectrum of the elementary fields. The masses of the massive gauge bosons are given by
\begin{align}
 \mA^{2} = \frac{1}{2} \frac{N}{P(N-P)} g^{2}w^{2}.
\end{align}
Correspondingly, we obtain $2P(N-P)$ would-be Goldstone modes which are not present in the elementary spectrum as they are BRST non-singlets and $N^{2}-1-2P(N-P)$ real scalar degrees of freedom which are the Higgs fields. 
They correspond to the unbroken generators and have mass $m \geq 0$. Moreover, we can already predict that $P^{2}-1$ of these scalar fields have the same mass denoted by $m_{\mathrm{P}}$ and $(N-P)^{2}-1$ masses $m_{\mathrm{N}-\mathrm{P}}$ are degenerated as well due to the group theoretical structure. These scalar fields belong to the $\SU(P)$ and $\SU(N-P)$ subgroups, respectively. 
Finally, there is the massive Higgs field corresponding to the generator of the vev and thus to the invariant $\U(1)$ subgroup. We denote its mass by $\mH$ in the following.
These masses read\footnote{Note some particularities. For sufficiently large $N$ and $N-1>P>2N/3$, the potential obeys only a saddle point as the condition that $\Sigma^{0}$ exhibits at most two different eigenvalues is a necessary but not sufficient condition that the potential has an extremum in that direction.  For $P=N-1$, a minimum can only be obtained in case $\tilde{\lambda}<0$. In this case there are no scalar fields with mass $m_{\mathrm{N}-\mathrm{P}}$.}
\begin{align}
 \mH^{2} &= \lambda w^{2} + 2\frac{(N-P)^{3}+P^3}{N^2P(N-P)}\tilde{\lambda}w^{2}, \label{eq:adjmassscalar} \\
 m_{\mathrm{P}}^{2} &= \frac{2N-3P}{P(N-P)}\tilde{\lambda}w^{2}, \qquad 
 m_{\mathrm{N}-\mathrm{P}}^{2} = \frac{3P-N}{P(N-P)}\tilde{\lambda}w^{2}. \notag
\end{align}
Thus, the elementary mass spectrum of $\SU(N)$ gauge theories is more involved than in the fundamental case, especially for increasing $N$. 
We will discuss some specific examples and particularities of some models below. 
Moreover, the different physical theories given by the different gauge orbits influence the physical gauge-invariant spectrum.

\subsection{Gauge-invariant spectrum}
\label{sec:adjoint-gi}

Before, we turn to the breaking patterns of some example groups and their precise spectrum, we discuss some of the properties of the gauge-invariant spectrum in general.
Neglecting the cubic term in Eq.~\eqref{eq:potadj}, the global symmetry group is given by a discrete $\mathbb{Z}_{2}$ symmetry. Thus, we classify our states in $\mathbb{Z}_{2}$ even ($+$) and $\mathbb{Z}_{2}$ odd ($-$) states. The lightest $\mathbb{Z}_{2}$ odd state is again necessarily absolutely stable.

We start the discussion of the gauge-invariant spectrum with the $1^-$ channel. Inspired from the fundamental 
case, an operator with minimal field content which is $\mathbb{Z}_2$ even is given by $\tr[\Sigma D_\mu \Sigma]$. The simplest $\mathbb{Z}_2$ odd state is $\tr[\Sigma^2 D_\mu\Sigma]$.\footnote{The even simpler state $\tr[D_\mu\Sigma]$ vanishes identically.}

However, those states do not expand to an elementary gauge field $A_{i}^{\mu}$, since $\tr[\Sigma^{n} D_{\mu} \Sigma] = \tr[ \Sigma^{n} \pd_{\mu} \Sigma] = \pd_{\mu}\tr\Sigma^{n+1}/(n+1)$, but in leading order to an elementary scalar field. Of course, they contribute to the pole structure of the $1^{-}$ channel but do not give rise to a vector particle as they have a pole only in the longitudinal component due to the partial derivative. This is similar to the fundamental case where we also observed that the mass pole of the scalar fields appears in the gauge-invariant vector state.

Nevertheless, it is also possible to construct operators which expand to a single gauge field in leading order, following the SU(2) fundamental case in \cite{Frohlich:1981yi}. These (minimally non-local, i.\ e.\ build from locally gauge-invariant) operators read
\begin{align}
O^{\mu}_{1^{-}_-}(x) &= \frac{\pd_{\nu}}{\pd^{2}}~\tr\big[\, \Sigma F^{\mu\nu}\,\big](x),  \label{eq:vectorZ2even} \\
O^{\mu}_{1^{-}_+}(x) &= \frac{\pd_{\nu}}{\pd^{2}}~\tr \big[\,\Sigma^{2} F^{\mu\nu}\,\big](x),  \label{eq:vectorZ2odd}
\end{align}
for the $\mathbb{Z}_{2}$ odd and even state, respectively.\footnote{Note that similar operators can also be constructed in the fundamental case for an arbitrary $\SU(N)$ theory, e.g., given by $\frac{\pd_{\nu}}{\pd^{2}}(\phi^{\dagger}F^{\mu\nu}\phi)$. However, the operator \eqref{eq:fund1-} provided in Sec.~\ref{fundgi} has less field content and is sufficient to get a prediction for the ground state mass of  the vector channel. Of course, the operator in Eq.~\eqref{eq:fund1-} and the one constructed from the field strength tensor as well as others have to be investigated in a detailed spectroscopy of the states as they carry the same quantum numbers.}
These states expand in leading order in the FMS expansion and in leading order in the elementary gauge field to
\begin{align}
\begin{split}
O^{\mu}_{1^{-}_-}(x)  &= - w\, \tr\big[\, \Sigma^{0} A_{\perp}^{\mu}\,\big] + \mathcal{O}(A^{2},\sigma),  \\
O^{\mu}_{1^{-}_+}(x) &= - w\, \tr\big[\, (\Sigma^{0})^{2} A_{\perp}^{\mu}\,\big] + \mathcal{O}(A^{2},\sigma),
\end{split}
\label{eq:masslessVector}
\end{align}
where $A_{\perp}^{\mu} = (\delta^{\mu}_{\nu} - \pd^{\mu}\pd_{\nu}/\pd^{2})A^{\nu}$ is the transversal component of the gauge field. 
Both operators expand to massless gauge fields as only those gauge fields survive the trace which are associated to generators of the Cartan algebra and thus are diagonal. 
More precisely,  the right hand side for both operators in Eq.~\eqref{eq:masslessVector} is given by the elementary gauge field that corresponds to the unbroken $\U(1)$ group given by the generator $\Sigma^{0}$.\footnote{There is only one particular exception for the $\mathbb{Z}_{2}$ even state for even $N$ for the breaking pattern $\SU(N) \to \SU(N/2)\times \SU(N/2)\times \U(1)$, see the discussion at the end of this section. Nonetheless, the ground state of this operator remains massless as it is given by two propagating massless gauge fields.} 
Thus the FMS mechanism predicts two massless physical states in the gauge-invariant vector channel of an arbitrary $\SU(N>2)$ gauge theory with a Higgs in the adjoint representation and discrete $\mathbb{Z}_{2}$ symmetry. 
Note that for $\SU(2)$ all $\mathbb{Z}_{2}$ even operators vanish identically in the vector channel and thus for $\SU(2)$ we get only a single massless vector state, see also the discussion in Sec.~\ref{sec:SU(2)adjoint}.

The results of Eq.~\eqref{eq:masslessVector} are a prediction of massless composite vector states. Moreover, this appears at the current time within the gauge-invariant setting to create massless vector states as physically observable particles. This is particularly interesting in the setting of grand-unified theories, where a U(1) with a massless vector particle has to be 'broken out' of a non-Abelian gauge theory \cite{Langacker:1980js,Bohm:2001yx}.

Further, we would like to emphasize that also a term quadratic in the gauge fields occurs to leading order in the FMS expansion. However, these as well as the higher order terms from the FMS expansion and the elementary scalar contribution from the operator $\tr[\Sigma^{n} D_{\mu} \Sigma]$ will not alter the ground state pole structure as these terms will correspond to scattering states or resonances in the $1^{-}$ channel or might even be massless, depending on the precise parameters and gauge group. Thus, the situation is different from the fundamental case, where the ground state mass of the $1^{-}$ channel is either $\MA$ or $\mh$. Here, we predict always two massless vector states.

In the $0^{+}$ channel, the simplest possible (minimal field content) operator is given by
\begin{align}
 O_{0^{+}_+}(x)  = \tr\big[\, \Sigma^{2} \,\big](x). 
 \label{eq:scalarZ2even}
\end{align}
This operator expands in the FMS description as
\begin{align}
 O_{0^{+}_+}(x)  =  \frac{w^{2}}{2} + w H(x) + \frac{1}{2} \sigma_{i}(x)\sigma_{i}(x).
\end{align}
Thus, we obtain in leading order the Higgs excitation $H(x) = \Sigma^{0}_{i}\sigma_{i}(x)$.
However, the situation is more subtle than in the fundamental case, as the term  $\sigma_{i}(x)\sigma_{i}(x)$ does not only contain the product of two massive Higgs fields $H(x)$ but also the scalar degrees of freedom belonging to the unbroken generators forming the invariant $\SU(P)$ and $\SU(N-P)$ subgroups which are not would-be Goldstone bosons and therefore present in the elementary spectrum. Whether these are massive or even massless depends on the details of the considered theory. 
In most cases, they have a mass given by Eq.~\eqref{eq:adjmassscalar}.
For some particular theories, however, they can be massless, e.g., see the discussion in Sec.~\ref{sec:SU(3)adjoint} or consider the case $P=2N/3$ or $P=N/3$ in Eq.~\eqref{eq:adjmassscalar}.

Therefore, the correlator $\langle O_{0^{+}_+}(x)^\dagger O_{0^{+}_+}(y) \rangle$ contains also the propagation of two degrees of freedom with mass $m_{\mathrm{P}}$ from position $x$ to position $y$ as well as two degrees of freedom with mass $m_{\mathrm{N}-\mathrm{P}}$ which is a bound state operator with mass $2m_{\mathrm{P}}$ or $2m_{\mathrm{N}-\mathrm{P}}$, respectively. 
We can suppose $N/2 \leq P < N$ without loss of generality.  Then, $0 \leq m_{\mathrm{P}} \leq m_{\mathrm{N}-\mathrm{P}}$. 
In case $\mH < 2m_{\mathrm{P}}$ the ground state mass is given by the mass of the scalar field radial to the direction of $\Sigma^{0}_{i}$. However, in case $\mH > 2m_{\mathrm{P}}$ the ground state mass is given by $2m_{\mathrm{P}}$ and an excited state is predicted at mass $\mH$ as well as a trivial scattering state with twice this mass. Which scenario is realized depends on the ratio of the couplings $\lambda$ and $\tilde{\lambda}$ as well as on the gauge group and the breaking pattern characterized by the numbers $N$ and $P$, respectively.

Moreover, we have the $\mathbb{Z}_{2}$ odd state in the $0^{+}$ channel given by
\begin{align}
 O_{0^{+}_-}(x)  = \tr\big[\, \Sigma^{3} \,\big](x).
 \label{eq:sigma3}
\end{align}
This operator can show distinct results for the different physical phases of the theory. Similar to the considerations above, this state can expand to elementary states which are massless or massive. 
We exemplify its different realizations by the following investigation. 

The FMS expansion of the scalar $\mathbb{Z}_{2}$ odd operator is given by
\begin{align}
 O_{0^{+}_-} = w^{3}\, \tr\, {\Sigma^{0}}^{3} + 3w^{2}\, \tr ({\Sigma^{0}}^{2}\sigma) + 3w\, \tr (\Sigma^{0}\sigma^{2}) + \tr\, \sigma^{3}.
\end{align}
To analyze its spectrum, it is convenient to perform a basis change of the generators of the Cartan subalgebra, i.e., a field redefinition of the fields in the Cartan. We leave the first $P-1$ elements of the Cartan algebra unchanged, i.e., they are given by the first $P-1$ diagonal matrices of the generalized Gell-Man matrices with rank two to $P$. For the Cartan elements $P$ to $N-2$, we choose the following block-diagonal matrices, $T = \begin{pmatrix} 0 & 0 \\ 0 & t \end{pmatrix}$, with $t$ one of the $N-P-1$ Cartan generators of the $N-P$ subgroup. For the remaining generator, we use the actual direction of the vev $\Sigma_{0}$, which can be parameterized as
\begin{align}
 \Sigma^{0} = \frac{1}{2} \sqrt{\frac{2}{NP(N-P)}} \begin{pmatrix} (N-P) \mathbbm{1}_{P} & 0 \\ 0 & -P \mathbbm{1}_{N-P} \end{pmatrix}
\end{align}
with $\mathbbm{1}_{x}$ the $x \times x$ unit matrix.

Now, it is straightforward to check that the leading order contribution in the fluctuating field is only given by the Higgs excitation associated to the vev, $\tr ({\Sigma^{0}}^{2}\sigma) = H \tr ({\Sigma^{0}}^{3})$. Thus, we generically get a pole at $\mH$ except for two cases. The first exception manifests for $\SU(3)$ and is discussed in Sec.~\ref{sec:SU(3)adjoint}. The second is given for even $N$ and the breaking pattern $\SU(N) \to \SU(N/2) \times \SU(N/2) \times \U(1)$. For this particular choice of the vev, $\tr ({\Sigma^{0}}^{2}\sigma)$ vanishes identically as ${\Sigma^{0}}^{2} {\sim} \mathbbm{1}$. 
At next to leading order, $\tr (\Sigma^{0}\sigma^{2})$, we obtain two fluctuating elementary fields propagating from $x$ to $y$. For the latter breaking pattern, these will make up the ground state mass for the $\mathbb{Z}_{2}$ odd operator. For all other theories, it depends again on the precise couplings as to whether $\mH$ is smaller than $2m_{\mathrm{P}}$ (or $2m_{\mathrm{N}-\mathrm{P}}$) and defines the ground state mass, similar to the $\mathbb{Z}_{2}$ even operator.

In addition to the predicted ground state masses of both scalar operators at tree level, these two operators can also have various additional excitations. Suppose, $\mH<2m_{\mathrm{P}}<2m_{\mathrm{N}-\mathrm{P}}<2\mH$. Then, the ground state mass is given by $\mH$ and each operator has a next-level state with mass $2m_{\mathrm{P}}$ and a next-to-next-level state with mass $2m_{\mathrm{N}-\mathrm{P}}$. All these states are either nontrivial scattering states or resonances as every state in the scalar channel can decay to at least two of the massless ground states in the vector channel.
Of course, similar conclusions hold for $2m_{\mathrm{P}}<\mH$ or other mass ratios.

Thus, the adjoint case has a much broader variety in the spectrum than the fundamental case. Depending on the physical realization of the theory, also see App.~\ref{a:state}, such a theory can have different numbers of observable states.

\subsection{SU(2) gauge theory}
\label{sec:SU(2)adjoint}
After these general considerations on the spectra of gauge theories with an adjoint Higgs field, we will now discuss some example theories to illustrate the different spectra for different realizations of the physical theories as well as some particularities of $\SU(2)$ and $\SU(3)$.

We start with the almost trivial example $\SU(2)$. The only nontrivial breaking pattern is $\SU(2) \to \U(1)$. The cubic term of the potential vanishes identically and the quartic $\tr \Sigma^{4}$ can be written as $(\tr \Sigma^{2})^{2}$, as $\tr \Sigma^{2}$ is the only invariant Casimir of $\SU(2)$. Thus, we set $\tilde\lambda = 0$ without loss of generality. 

The elementary spectrum is given by one massive Higgs excitation with mass $\mH^{2} = \lambda w^{2}$, two massive gauge bosons with mass $\mA^{2} = g^{2} w^{2}$ and one massless gauge boson. A convenient choice for the vev is $\Sigma^{0}_{i} = \delta_{{i3}}$.

Considering the gauge-invariant bound-state spectrum, we first note that all $\mathbb{Z}_{2}$ odd operators in the scalar channel, $\tr(\Sigma^{2n+1})$, and $\mathbb{Z}_{2}$ even operators in the vector channel, $\tr(\Sigma^{2n} F^{\mu\nu})$, vanish identically due to the properties of the Pauli matrices. Thus, we have only one state within the $0^{+}$ as well as $1^{-}$ channel, modulo higher excitations. For the scalar channel, the operator \eqref{eq:scalarZ2even} expands to, 
\begin{align}
 O_{0^{+}_+}(x)  =  \frac{w^{2}}{2} + w H(x) + \frac{1}{2} H^{2}(x),
\end{align}
where we have omitted the contributions from the would-be Goldstone modes, given by $\sigma_{i}\Sigma^{0}_{i}=0$ in this specific model, as these cancel anyway once the physical spectrum is considered. Thus, we get a pole at the mass of the elementary Higgs field as well as a trivial scattering state at twice its mass, similar to the fundamental case. 

For the vector channel, we predict generically a massless state as the ground state for the operator \eqref{eq:vectorZ2even}. This is also seen in a nonperturbative lattice investigation \cite{Lee:1985yi}. Going beyond  the leading order contribution in the fluctuating fields, we also predict a next-level state, being either a nontrivial scattering state or a resonance, at twice the mass of the elementary massive gauge fields. A brief summary of the spectra can be found in Tab.~\ref{tab:su2ad}.

\begin{table}[t!]
\begin{center}
\begin{tabular}{c|ccc|ccccc}

\multicolumn{4}{c}{\quad elementary spect.} & \multicolumn{5}{c}{gauge-invariant spect.} \cr

\toprule

$J^P$ & Field & Mass & Deg. & $\mathbb{Z}_{2}$ & Op. & Mass & Next-level state & Deg.\cr
\hline

$0^+$ & $H$ & $\mH$ & $1$ & $+$ & $O_{0^+_+}$ & $\mH$ & - & $1$\cr
\hline
$1^-$ & $A^\mu_{3}$ & $0$ & $1$ & $-$ & $O^\mu_{1^-_{-}}$ & $0$ & $2\mA$ & $1$\cr
 & $A^\mu_{1,2}$ & $\mA$ & $2$ &  &  &  & \cr

\toprule
\end{tabular}
\caption{Left: Gauge-variant spectrum of an $\SU(2)$ gauge theory with a single scalar field in the adjoint representation. Right: Gauge-invariant (physical) spectrum of the theory. $\mH = \sqrt{\lambda} w$ denotes the mass of the Higgs, $\mA = g w$ is the mass of the two charged elementary gauge bosons. The next-level state of the vector channel will always be a scattering state or a resonance. The definitions of the fields and operators can be found in the main text.}
\label{tab:su2ad}

\end{center}
\end{table}

\subsection{SU(3) gauge theory}
\label{sec:SU(3)adjoint}
The next example is the $\SU(3)$ case, which has a variety of new features compared to the $\SU(2)$ case. Most important, the $\mathbb{Z}_{2}$ odd and even operators are nonvanishing in the scalar and vector channel, respectively. Thus, new states, i.e., observable particles, are present. In addition, we get two different breaking patterns and a second invariant Casimir can be constructed, $\tr \Sigma^{3}$. So further properties of the spectrum for the adjoint case can be exemplified but also particularities which are only present in this model.


\begin{table*}[t!]
\begin{center}
\begin{tabular}{c|ccc|ccccc}

\multicolumn{4}{r}{\quad elementary spectrum} & \multicolumn{4}{c}{\quad gauge-invariant spectrum } \cr
\toprule

$J^P$ & Field & Mass & Deg. & $\mathbb{Z}_{2}$ & Op. & Mass & Next-level state & Deg.\cr
\hline

$0^+$ & $\sigma_{3}\equiv H$ & $\mH$ & $1$ & $+$ & $O_{0^+_+}$ & $0$ & $\mH$ & $1$\cr
 & $\sigma_{8}$ & $0$ & $1$ & $-$ & $O_{0^+_-}$ & 0 & $2\mH$ & $1$\cr
\hline
$1^-$ & $A^\mu_{3,8}$ & $0$ & $2$ & $+$ & $O^\mu_{1^-_{+}}$ & $0$ & $2m_{\mathrm{A}_{4}}$  & $1$\cr
 & $A^\mu_{1,2}$ & $m_{\mathrm{A}_1}$ & $2$ & $-$ & $O^\mu_{1^-_{-}}$ & $0$ & $2m_{\mathrm{A}_{4}}$  & $1$\cr
 & $A^\mu_{4,5,6,7}$ & $m_{\mathrm{A}_4}$ & $4$ & \cr

\toprule
\multicolumn{8}{l}{breaking pattern: $\SU(3) \to \U(1) \times \U(1)$ ($\Sigma^{0}_{3}=1$)}
\end{tabular}
\hfill
\begin{tabular}{c|ccc|ccccc}

\multicolumn{4}{r}{\quad elementary spectrum} & \multicolumn{4}{c}{\quad gauge-invariant spectrum } \cr

\toprule

$J^P$ & \quad Field \quad & Mass & Deg. & $\mathbb{Z}_{2}$ & Op. & Mass & Next-level state & Deg.\cr
\hline

$0^+$ & $\sigma_{1,2,3}$ & $0$ & $3$ & $+$ & $O_{0^+_+}$ & $0$ & $\mH$ & $1$\cr
 & $\sigma_{8}\equiv H$ & $\mH$ & $1$ & $-$ & $O_{0^+_-}$ & 0 & $\mH$ & $1$\cr
\hline
$1^-$ & $A^\mu_{1,2,3,8}$ & $0$ & $4$ & $+$ & $O^\mu_{1^-_{+}}$ & $0$ & $2\mA$  & $1$\cr
 & $A^\mu_{4,5,6,7}$ & $\mA$ & $4$ & $-$ & $O^\mu_{1^-_{-}}$ & $0$ & $2\mA$ & $1$\cr
 & & & & &  \cr
\toprule
\multicolumn{8}{l}{breaking pattern: $\SU(3) \to \SU(2) \times \U(1)$ ($\Sigma^{0}_{8}=1$)}
\end{tabular}

\caption{Gauge-variant and gauge-invariant spectrum of an $\SU(3)$ gauge theory with a single scalar field in the adjoint representation with discrete $\mathbb{Z}_{2}$ symmetry for the two different breaking patterns. Left table: The breaking pattern is given by $\SU(3) \to \U(1) \times \U(1)$ and the gauge orbit is characterized by $\Sigma^{0}_{3}=1$, $\Sigma^{0}_{8}=0$. The masses are given by $\mH = \sqrt{\lambda} w$, $m_{\mathrm{A}_1} = gw$, and $m_{\mathrm{A}_4} = gw/2$. Right table: The breaking pattern is given by $\SU(3) \to \SU(2) \times \U(1)$ and the gauge orbit is characterized by $\Sigma^{0}_{3}=0$, $\Sigma^{0}_{8}=1$. The mass $\mA = \sqrt{3}gw/2$.}
\label{tab:su3ad}

\end{center}
\end{table*}


Although $\tr \Sigma^{3} \neq 0$, we first demand a $\mathbb{Z}_{2}$ invariant Lagrangian in order to analyze the spectrum. Thus, we set $\gamma=0$ in the scalar potential~\eqref{eq:potadj} as well as $\tilde\lambda = 0$ without loss of generality in analogy to the $\SU(2)$ case. Note, that the pure scalar part of the action has an enhanced $\mathrm{O}(8)$ symmetry in this case as the potential is only build up from the invariant $\tr \Sigma^{2} = \Sigma_{i}\Sigma_{i}/2$. 

The two breaking patterns of $\SU(3)$ are given by $\SU(2)\times \U(1)$ and $\U(1)\times \U(1)$. 
Parameterizing all possible vevs by $\Sigma^{0} = \Sigma^{0}_{3} T_{3} + \Sigma^{0}_{8} T_{8}$ with $(\Sigma^{0}_{3})^{2} + (\Sigma^{0}_{8})^{2} = 1$, the former breaking pattern can be realized by the three particular combinations $\big(0,1\big)$, $\big(\sqrt{3}/2,1/2\big)$, and $\big(\sqrt{3}/2,-1/2\big)$ regarding the tuple $(\Sigma^{0}_{3},\Sigma^{0}_{8})$. All other combinations result in an invariant $\U(1)\times\U(1)$ subgroup.
Note that also the latter breaking pattern minimizes the potential due to the enhanced symmetry of the scalar potential even if $\Sigma^{0}$ has three different eigenvalues. This is a particularity of $\SU(3)$ with a discrete $\mathbb{Z}_{2}$ symmetry and will be no longer the case once a nonvanishing $\gamma$ in the potential is allowed or larger gauge groups are considered where $\tr \Sigma^{4}$ is another invariant Casimir for $N>3$.

While the spectrum of the elementary vector states differs for both breaking patterns, the gauge invariant spectrum of the vector states remains the same. 
The mass spectrum of the elementary gauge bosons reads
\begin{align*}
 &(\MA^{2}) = g^{2}w^{2} \diag \left(  {\Sigma^{0}_{3} }^{2}, {\Sigma^{0}_{3} }^{2}, 0,  \frac{(\Sigma^{0}_{3}  + \sqrt{3}\Sigma^{0}_{8})^2}{4}, \right.  \\ 
 &\qquad \left. \frac{(\Sigma^{0}_{3}  + \sqrt{3}\Sigma^{0}_{8})^2}{4}, \frac{(\Sigma^{0}_{3}  - \sqrt{3}\Sigma^{0}_{8})^2}{4}, \frac{(\Sigma^{0}_{3}  - \sqrt{3}\Sigma^{0}_{8})^2}{4}, 0\right).
\end{align*}
Thus in general, the theory contains at least two massless gauge bosons corresponding to the two generators of the Cartan subalgebra and at most six massive fields, which we can group in three pairs where the masses of the gauge bosons of each pair are degenerate. 
In case of a breaking to $\SU(2)\times \U(1)$, e.g., by choosing $\Sigma^{0}_{3}=0$ and $\Sigma^{0}_{8}=1$, we obtain four massless gauge bosons corresponding to the invariant subgroup and four degenerated massive gauge bosons with mass $\mA = \sqrt{3} g w/2$. 
For completeness, there are also three particular directions, $\big( 1,0 \big)$, $\big( 1/2,\sqrt{3}/2 \big)$, and $\big( 1/2,-\sqrt{3}/2 \big)$, with little group $\U(1)\times\U(1)$, for which two of the mass pairs further degenerate such that the spectrum contains two massive vector bosons with mass $m_{\mathrm{A}_{1}} = g w$ and four degenerated fields with mass $m_{\mathrm{A}_{4}} = g w/2$.

Even though the spectrum of the elementary vector states is different for the different physical realizations of the theory characterized by the different directions of the vev in the Cartan algebra, the gauge-invariant spectrum always contains two massless vector states distinguished by their global $\mathbb{Z}_{2}$ parity as described in Sec.~\ref{sec:adjoint-gi}.

For the scalar channel, the elementary spectrum is much clearer. First, we have four or six would-be Goldstone bosons, depending on the respective breaking pattern, which mix with the longitudinal parts of the massive gauge bosons. Second, there is always one massive Higgs field with mass $\mH = \sqrt{\lambda}w$, corresponding to the generator proportional to the vev, $H=\Sigma^{0}_{i}\sigma_{i}$.  The remaining three or one scalar field(s) belong to the elementary spectrum of the theory and are Higgs fields as well.
They are massless for the $\SU(3)$ case with a discrete $\mathbb{Z}_{2}$ symmetry imposed on the scalar potential. 
We emphasize at this point that they are not Goldstone bosons as they belong to the remaining unbroken generators of the gauge group, $[T_{i},\Sigma^{0}]=0$. The fact that they are massless is an accident due to the enhanced $\mathrm{O}(8)$ symmetry of the scalar potential.

The presence of massless fields in the elementary spectrum has a direct impact on the gauge-invariant spectrum of the composite operators \eqref{eq:scalarZ2even} and \eqref{eq:sigma3}. Although, the $\mathbb{Z}_{2}$ even operator \eqref{eq:scalarZ2even} expands in leading order to the massive elementary Higgs field $H$, it also contains the propagation of two massless scalar degrees of freedom at next-to-leading order in the FMS expansion. Thus, the FMS mechanism predicts at least at tree level a massless bound state for this operator.\footnote{Of course, it is possible that also this state, as well as other massless bound states, acquires a non-vanishing mass due to quantum corrections, e.g., in analogy to glueballs.} Similar conclusions can be drawn for the $\mathbb{Z}_{2}$ odd operator. Depending on the actual direction of the vev in the Cartan space, either the leading order or next-to-leading order expands to massless scalar fields. For instance, the leading order expansion contains the massless field $\sigma_{8}$ for $\Sigma^{0} = T_{3}$ and thus the ground state mass is $m_{0^{+}_{-}} = 0$. The next-to leading order in the FMS expansion predicts the mass of the next-level state which is $2\mH$. For $\Sigma^{0} = T_{8}$ the leading order in the FMS expansion contains also only $\sigma_{8}$ but which is this time massive. However, the next to-leading-order contribution $\mathcal{O}(\sigma^{2})$ contains products of the massless fields $\sigma_{1,2,3}$ such that the 2-point function of this operator has a pole at vanishing mass on the right hand side. A summary of these two example breaking patterns can be found in Tab.~\ref{tab:su3ad}.


\begin{table}[t!]
\begin{center}
\begin{tabular}{c|ccc|cccc}

\multicolumn{4}{r}{\quad elementary spectrum} & \multicolumn{4}{c}{\quad gauge-invariant spectrum } \cr
\toprule

$J^P$ & Field & Mass & Deg. & Op. & Mass & Next-level state & Deg.\cr
\hline

$0^+$ & $H$ & $\mH$ & $1$ & $O_{0^+}$ & $\mH$ & $2m_{\mathrm{P}}$ & $1$\cr
 & $\sigma_{1,2,3}$ & $m_{\mathrm{P}=2}$ & 3 & \cr
\hline
$1^-$ & $A^\mu_{1,2,3,8}$ & $0$ & $4$ & $O^\mu_{1^-}$ & $0$ & $2m_{\mathrm{A}}$   & $1$\cr
 & $A^\mu_{4,5,6,7}$ & $\mA$ & $4$ &  \cr

\toprule
\multicolumn{8}{l}{breaking pattern: $\SU(3) \to \SU(2) \times \U(1)$} 
\end{tabular}

\caption{Gauge-variant and gauge-invariant spectrum of an $\SU(3)$ gauge theory with a single scalar field in the adjoint representation. The scalar potential exhibits an explicit $\mathbb{Z}_{2}$ breaking term, $\gamma\, \tr\Sigma^{3}$. Thus, there is no $\mathbb{Z}_{2}$ classification of the gauge-invariant spectrum. We have assumed $\mH<2m_{\mathrm{P}=2}$ for the ground state mass of the scalar channel otherwise the ground state mass would be $2m_{\mathrm{P}}$ and the next-level state has mass $\mH$.}
\label{tab:su3ad2}

\end{center}
\end{table}


So far, we only investigated the spectrum for theories with a discrete $\mathbb{Z}_{2}$ symmetry for the scalar field. Allowing for an explicit breaking term will not only influence the gauge-invariant spectrum indirectly as the elementary spectrum is changed. It also has a direct impact as transitions between the $\mathbb{Z}_{2}$ even and odd states are allowed. Therefore, the number of observable gauge-invariant states is altered as less different channels exist. A straightforward example for this is the vector channel of the theory. The former two massless states distinguished by their $\mathbb{Z}_{2}$ quantum number are now in the same $1^{-}$ channel and thus have overlap with the same ground state. Thus, the FMS description predicts only one massless state for any $\SU(N)$ gauge theory with an adjoint Higgs once the Lagrangian does not exhibit a $\mathbb{Z}_{2}$ symmetry. 

The same statement can be formulated for the scalar channel. In addition, the situation is more involved for $\gamma \neq 0$ in the scalar potential given in Eq.~\eqref{eq:potadj}. In this case, the potential has minima only in the three equivalent directions which lead to a breaking pattern $\SU(3)\to\SU(2)\times\U(1)$ according to the lemma of \cite{Ruegg:1980gf}. The elementary scalar spectrum has a Higgs state with mass $\mH^{2} = \lambda w^{2} - \frac{\sqrt{3}}{2}\gamma w$ and three degenerate Higgs states with mass $m_{\mathrm{P}=2}^{2} = \frac{\sqrt{27}}{2}\gamma w$. Thus, the gauge-invariant ground state in the scalar channel is now massive and either $\mH$ for $2 \lambda w < \sqrt{75}\gamma$ or $2m_{\mathrm{P}=2}$ for $2 \lambda w > \sqrt{75}\gamma$. A brief summary of the spectrum is given in Tab.~\ref{tab:su3ad2} for the former case.

\subsection{SU(4) gauge theory}
Finally, we discuss the spectrum of $\SU(4)$. This is the smallest group for which we have the full set of invariant Casimirs regarding perturbative renormalizability and we are able to directly apply the formulas provided in Sec.~\ref{sec:adjoint-gv}. 
For simplicity, we again impose a discrete $\mathbb{Z}_{2}$ symmetry on the Lagrangian, i.e., $\gamma=0$. The nonvanishing coupling $\tilde\lambda\, \tr\Sigma^{4}$ will induce nonvanishing masses for the elementary scalar fields which belong to the unbroken generators of the theory but are not the Higgs excitation radial to $\Sigma^{0}_{i}$. Even if $\tilde\lambda$ is zero at the classical level, it will be generated due to quantum corrections from the gauge bosons as it is not protected by a symmetry.
The scalar potential has minima only for the breaking patterns $\SU(4)\to \SU(3)\times \U(1)$ and $\SU(4) \to \SU(2)\times \SU(2)\times \U(1)$. 
While the latter corresponds to the global minimum for $\tilde\lambda >0$, the former breaking pattern is favored for $\tilde\lambda <0$.
The other directions in the Cartan space associated to little groups $\SU(2)\times \U(1)^{2}$ and $\U(1)^3$ are not extrema of the potential and can be ignored.


\begin{table}[t!]
\begin{center}
\begin{tabular}{c|ccc|ccccc}

\multicolumn{4}{r}{\quad elementary spectrum} & \multicolumn{4}{c}{\quad gauge-invariant spectrum } \cr
\toprule

$J^P$ & Field & Mass & Deg. & $\mathbb{Z}_{2}$ & Op. & Mass & Next-level state & Deg.\cr
\hline

$0^+$ & $H$ & $\mH$ & $1$ & $+$ & $O_{0^+_+}$ & $\mH$ & $2m_{\mathrm{P}=3}$ & $1$\cr
 & $\sigma_{1,\dots,8}$ & $m_{\mathrm{P}=3}$ & $8$ & $-$ & $O_{0^+_-}$ & $\mH$ & $2m_{\mathrm{P}=3}$ & $1$\cr
\hline
$1^-$ & $A^\mu_{15}$ & $0$ & $2$ & $+$ & $O^\mu_{1^-_{+}}$ & $0$ & $2\mA$  & $1$\cr
 & $A^\mu_{1,\dots,8}$ & $0$ & $8$ & $-$ & $O^\mu_{1^-_{-}}$ & $0$ & $2\mA$ & $1$\cr
 & $A^\mu_{9,\dots,14}$ & $\mA$ & $6$ & \cr

\toprule
\end{tabular}

\caption{Gauge-variant and gauge-invariant spectrum of an $\SU(4)$ gauge theory with a single scalar field in the adjoint representation with discrete $\mathbb{Z}_{2}$ symmetry for the breaking pattern $\SU(4) \to \SU(3) \times \U(1)$, e.g., realized by $\Sigma^{0}=T_{15}$. We assume $\mH<2m_{\mathrm{P}=3}$ for the gauge invariant spectrum, here.} 
\label{tab:su4ad}

\end{center}
\end{table}



\begin{table}[t!]
\begin{center}

\begin{tabular}{c|ccc|ccccc}

\multicolumn{4}{r}{\quad elementary spectrum} & \multicolumn{4}{c}{\quad gauge-invariant spectrum } \cr

\toprule

$J^P$ & \quad Field \quad & Mass & Deg. & $\mathbb{Z}_{2}$ & Op. & Mass & N.-l. state & Deg.\cr
\hline

$0^+$ & $H$ & $\mH$ & $1$ & $+$ & $O_{0^+_+}$ & $\mH$ & $2m_{\mathrm{P}=2}$ & $1$\cr
 & $\sigma_{\SU(2)}$ & $m_{\mathrm{P}=2}$ & $3$ & $-$ & $O_{0^+_-}$ & $2m_{\mathrm{P}=2}$ & $2\mH$ & $1$\cr
 & $\bar\sigma_{\SU(2)}$ & $m_{\mathrm{P}=2}$ & $3$ & \cr
\hline
$1^-$ & $A^\mu_{\U(1)}$ & $0$ & $1$ & $+$ & $O^\mu_{1^-_{+}}$ & $0$ & $2\mA$  & $1$\cr
 & $A^\mu_{\SU(2)}$ & $0$ & $3$ & $-$ & $O^\mu_{1^-_{-}}$ & $0$ & $2\mA$ & $1$\cr
 & $A^\mu_{\SU(2)}$ & $0$ & $3$ & \cr
 & $A^\mu_{4,\dots,7,9,\dots,12}$ & $\mA$ & $8$ & \cr
\toprule
\end{tabular}

\caption{Gauge-variant and gauge-invariant spectrum of an $\SU(4)$ gauge theory with a single scalar field in the adjoint representation with discrete $\mathbb{Z}_{2}$ symmetry for the breaking pattern $\SU(4) \to \SU(2)^{2} \times \U(1)$, e.g., realized by $\Sigma^{0}= \sqrt{2/3}\,T_{8}+ \sqrt{1/3}\,T_{15}$. We assume $\mH<2m_{\mathrm{P}=2}$ for the gauge invariant spectrum, here. The field $\sigma_{\SU(2)}$ encodes the fields $\sigma_{1,2,3}$ which belong to the first remaining $\SU(2)$ group, while $\bar\sigma_{\SU(2)}$ encodes the three scalar fields which belong to the other remaining $\SU(2)$ group given by $\sigma_{13}$, $\sigma_{14}$, and $-\sqrt{1/3}\,\sigma_{8}+ \sqrt{2/3}\,\sigma_{15}$, for the exemplified $\Sigma^{0}$. Similar considerations hold for the gauge bosons.}  
\label{tab:su4ad2}

\end{center}
\end{table}


For the breaking pattern $\SU(4)\to \SU(3)\times \U(1)$ the elementary spectrum is given by nine massless gauge fields and six massive ones with mass $\mA^{2} = \frac{2}{3}g^{2}w^{2}$. Moreover, we obtain a massive Higgs with mass $\mH^{2} = (6\lambda + 7\tilde\lambda)w^{2}/6$ and eight degenerate scalar states invariant under the remaining $\SU(3)$ subgroup with mass $m_{\mathrm{P}=3}^{2} = -\tilde\lambda w^{2}/3$. In order that the scalar potential exhibits a minimum, the couplings have to fulfill the relation $6\lambda>-7\tilde\lambda$ and $\tilde\lambda<0$. The ground state mass created by the gauge-invariant $0^{+}_{+}$ or $0^{+}_{-}$ operator is given by either $\mH$ or $2m_{\mathrm{P}}$, depending on the ratios of the coupling constants. In the vector channel two massless vector bosons are predicted as discussed in detail above. 
The spectra are summarized in Tab.~\ref{tab:su4ad}.

For the breaking pattern $\SU(4)\to \SU(2)\times\SU(2)\times \U(1)$ the elementary spectrum is given by seven massless gauge fields and eight massive ones with mass $\mA^{2} = g^{2}w^{2}/2$. In the scalar sector, we have a massive Higgs, $\mH^{2} = (\lambda + \tilde\lambda/2)w^{2}$, three massive scalar fields invariant under the first $\SU(2)$ subgroup with mass $m_{\mathrm{P}=2}^{2} = \tilde\lambda w^{2}/2$ and three scalar fields invariant with respect to the second $\SU(2)$ subgroup with the same mass $m_{\mathrm{N}-\mathrm{P}=2}^{2} = \tilde\lambda w^{2}/2$. The gauge invariant spectrum for the scalar $\mathbb{Z}_{2}$ even operator depends on the ratio of the two quartic couplings. For $2\lambda < 3\tilde{\lambda}$, the ground state mass is given by $\mH$ and we obtain a further next-level state with mass $2m_{\mathrm{P}}$ at tree level and vice versa for $2\lambda > 3\tilde{\lambda}$. The ground state of the $\mathbb{Z}_{2}$ odd operator is generically given by $2m_{\mathrm{P}=2}$ for positive quartic couplings. The only possibility to obtain a ground state mass of $2\mH$ for this operator is given by $\lambda<0$. In this case $2\lambda + \tilde\lambda > 0$ has to hold to have a minimum as well as a potential bounded from below. For a brief summary of the gauge-variant and gauge-invariant spectrum for this particular breaking pattern see Tab.~\ref{tab:su4ad2}.

\section{SU(5) GUT as a toy model}
\label{sec:gut}

Moving away from the basic ingredients, we consider in this section the scalar-gauge sector of the prototype of grand unified theories, namely an $\SU(5)$ gauge theory with one scalar field in the adjoint representation $\Sigma$ and one in the fundamental representation $\phi$, but ignore the fermionic sector for simplicity \cite{Georgi:1974sy}. The Lagrangian has at least a $\U(1)$ symmetry acting on $\phi$. Depending on the precise form of the scalar potential, this symmetry can be enhanced to a $\mathbb{Z}_2\times \U(1)$ custodial symmetry in case it is invariant under the discrete transformation $\Sigma \to -\Sigma$.

\begin{table*}[t]
\begin{center}
\begin{tabular}{l|llc|ccccc}

\multicolumn{4}{c}{\quad elementary spectrum} & \multicolumn{4}{c}{\qquad gauge-invariant spectrum } \cr
\toprule

$J^P$ & Field & Mass & Degeneracy & $(\U(1),\mathbb{Z}_{2})$ & Operator & Mass & Next-level state & Degeneracy\cr
\hline

$0^+$ & $h$ & $\mh$ & $1$ & $(0,+)$ & $O_{0^{+}_{0+}}$ & $\mh$ & $\sim w$ & $1$ \cr
& $\varphi^{1,\dots,6}$ & $m_{\varphi^{1,\dots,6}}$ & $6$ & $(0,-)$ & $O_{0^{+}_{0-}}$ & $\mh$ & $\sim w$ & $1$ \cr
& $\sigma_{1,\dots,8}$ & $m_{\sigma_{1,\dots,8}}$ & $8$ & $(\pm 1,+)$ & $O_{0^{+}_{\pm 1+}}$ & $\sim w$ & $\sim w$ & $1$ \cr
 & $\sigma_{21,22,23}$ & $m_\sigma$ & $3$ & $(\pm 1,-)$ & $O_{0^{+}_{\pm 1-}}$ & $\sim w$ & $\sim w$ & $1$ \cr
 & $\sigma_{24}$ & $M_\sigma$ & $1$ & \cr
\hline
$1^-$ & $A^\mu$ & $\mA = 0$ & $1$ & $(0,+)$ & $O_{1^{-}_{0+}}$ & $\mA$ & $m_Z$ & $1$\cr
 & $W_{\pm}^\mu$ & $m_W$ & $2$ & $(0,-)$ & $O_{1^{-}_{0-}}$ & $\mA$ & $m_Z$ & $1$\cr
 & $Z^\mu$ & $m_Z$ & $1$ & $(\pm 1,+)$ & $O_{1^{-}_{\pm 1+}}$ & $\sim w$ & $\sim w$ & $1$\cr
 & $A_{9,\dots,14}^\mu$ & $m_L$ & $6$ & $(\pm 1,-)$ & $O_{1^{-}_{\pm 1-}}$ & $\sim w$ & $\sim w$ & $1$\cr
 & $A_{15,\dots,20}^\mu$ & $M_L$ & $6$ \cr

\toprule
\end{tabular}
\caption{Left: Gauge-variant spectrum of an $\SU(5)$ gauge theory with one scalar field in the fundamental representation and one in the adjoint representation. The fields listed here are all mass-eigenstates. The $8$ massless gluons which correspond to the unbroken $SU(3)$ gauge group are not listed.\\
Right: Gauge-invariant (physical) spectrum of the theory. In case the $\mathbb{Z}_{2}$ symmetry is explicitly broken, we obtain only one ground state within the different $\U(1)$ channels. For simplicity, we only indicate the order of magnitude for the masses of the heavy gauge invariant states.}
\label{tab:su5}
\end{center}
\end{table*}

\subsection{Gauge-variant description in a fixed gauge}
The perturbative construction works as follows: First the adjoint scalar acquires a vev $\langle\Sigma\rangle {\sim} T^{24}$, 
where $T^{24}$ is the following element of the Cartan subalgebra,
$T^{24}=\frac{1}{2}\sqrt{\frac{3}{5}}\text{diag}(-2/3,-2/3,-2/3,1,1)$. This choice of the vev of the scalar field in the adjoint representation breaks at some high scale, much larger than the electroweak scale, $\SU(5)$ to $\SU(3)_{\mathrm{C}}\times \SU(2)_{\mathrm{L}}\times \U(1)_{\mathrm{Y}}$ which is the standard model gauge group. 
The fundamental scalar acquires a vev $\langle\phi\rangle = v n/\sqrt{2}$, where $v$ is of the order of the electroweak scale. It is sufficient to consider a complex unit-vector of the form $n=(0,0,n^{3},0,n^{5})$ with $n^{5}\in \mathbb{R}$ as we are still allowed to perform appropriate $\SU(3)$ and $\SU(2)$ gauge transformations. In order to leave the strong interaction unbroken, it is important to impose constraints on the parameters of the scalar potential in such a way, that the potential energy of the scalar fields is minimized for a field configuration with $n=(0,0,0,0,1)$.
In this case, the fundamental Higgs field breaks the 
standard model group to $\SU(3)_{\mathrm{C}}\times \U(1)_{\mathrm{EM}}$. 
To achieve this, we can consider the following special realization of the Lagrangian~\eqref{lagrangian}
\begin{align}
{\cal L} =  &-\frac{1}{4} F_{i\,\mn}F^{\mn}_{i}  +  \tr [(D_{\m} \Sigma)^{\dagger} D^{\m} \Sigma] + 
(D_{\m} \phi)^{\dagger} (D^{\m} \phi) \notag \\ 
&-  V_\Sigma(\Sigma) - V_\phi(\phi^{\dagger}\phi) - \beta \phi^\dagger \Sigma^2\phi,
\label{eq:su5L}
\end{align}
where $V_\Sigma$ is defined in Eq.~\eqref{eq:potadj}.  The Lagrangian is invariant under the discrete $\mathbb{Z}_{2}$ transformation for $\gamma=0$. For nonvanishing $\gamma$ also the interaction term $\phi^{\dagger}\Sigma\phi$ has to be considered as this term will be at least generated by radiative corrections in case the $\mathbb{Z}_{2}$ symmetry is explicitly broken. The part of the potential which contains only the self interactions of the fundamental scalar $V_\phi$ is of the usual Mexican-hat form. The term proportional to $\beta$, mixing both scalar fields, is responsible for driving the system into the appropriate minimum where the strong interaction remains unbroken. A necessary condition is $\beta<0$. Moreover, we dropped the remaining possible perturbatively renormalizable term given by $\phi^{\dagger}\phi\, \tr\Sigma^{2}$ as this term leads only to an unimportant mass shift in the elementary scalar spectrum but does not influence the main results of the following section.

All together this leads to the following (perturbative) elementary particle spectrum: From the breaking of $\SU(5)$ to the standard model 
group we get $12$ heavy gauge bosons called leptoquarks and $12$ heavy Higgs bosons. From the breaking of the 
standard model group to $\SU(3)_C\times \U(1)_{EM}$ we obtain $3$ light gauge bosons, the $W^{\pm}$ and the $Z$, and 
$7$ additional Higgs bosons. Of course, there is fine tuning at work, such that only one of the $19$ Higgs bosons will be 
the standard model Higgs and the others will have masses of the order of the GUT scale denoted by $w$ in the following, as do the leptoquarks \cite{Bohm:2001yx}.

We want to point out that the last term in Eq.~\eqref{eq:su5L} gives rise to an off-diagonal mass matrix for the $\Sigma$ fields leading to a mixture of $\Sigma_{23}$ and $\Sigma_{24}$ depending on the parameters of the potential. Thus, a suitable field redefinition has to be performed such that a diagonal mass matrix is achieved. These redefined fields are used in the next subsection to identify the poles of the gauge-invariant states.

The gauge-variant spectrum of this theory is sketched on the left-hand side of Tab.~\ref{tab:su5}. The lightest scalar field with mass of the electroweak scale $\mh \sim v$ is denoted by $h$. We do not provide the explicit masses in terms of the masses of the elementary fields for all heavy gauge-invariant states for better readability. As these masses are of the order of the GUT scale, we simply indicate that they are ${\sim} w$.

\subsection{Gauge-invariant description}
We begin the analysis of the gauge-invariant spectrum for a Lagrangian with a $\mathbb{Z}_2\times \U(1)$  custodial symmetry. Therefore, we can group the states into 
$\U(1)$ singlets and non-singlets which can have $\mathbb{Z}_2$ even or odd parity.

Due to the increased field content several gauge-invariant operators can be constructed to analyze the spectrum of the different channels. For instance, some of the possible classes of operators for the scalar $\U(1)$-singlet $\mathbb{Z}_{2}$-even channel $0^{+}_{0+}$ are $(\phi^{\dagger}\phi)^{n}$, $\tr\,\Sigma^{2n}$, $(\phi^{\dagger}\Sigma^{2n}\phi)^{m}$, $\cdots$, with $n,m \in \mathbb{N}$. Analyzing these operators with the aid of the FMS prescription, we conclude that the ground state mass is given by the mass of the lightest fundamental perturbative Higgs $h$. Further, there are a number of excited states with mass of the order of the GUT scale $w$, e.g., with the mass of the fields $\sigma_{23}$ and $\sigma_{24}$ or twice the masses of the other elementary scalar fields. A similar spectrum appears for the $\U(1)$-singlet $\mathbb{Z}_{2}$-odd channel $0^{+}_{0-}$. The operator $\phi^{\dagger}\Sigma\phi$ can be used to straightforwardly compute that the mass of the ground state is again the mass of the lightest elementary fundamental Higgs excitation $h$. Thus, the gauge-invariant spectrum contains not only one but two scalar fields with mass of the order of the electroweak scale, while the elementary spectrum has only one light scalar field.

The situation becomes even more problematic once the vector channel is investigated. The vector $\U(1)$-singlet channel can be analyzed by the operators $\frac{\partial^\nu}{\partial^2}\tr[F_{\mu\nu}\Sigma^{2n}]$, $\I\phi^\dagger\Sigma^{2n}D_\mu\phi$, $\cdots$, or $\frac{\partial^\nu}{\partial^2}\tr[F_{\mu\nu}\Sigma^{2n+1}]$, $\I\phi^\dagger\Sigma^{2n+1}D_\mu\phi$, $\cdots$, for the $\mathbb{Z}_{2}$ even and odd case, respectively. These are inspired by the investigations in the previous sections but even more involved singlet operators can be studied, e.g., $\epsilon^{abcde}\epsilon^{\bar{a}\bar{b}\bar{c}\bar{d}\bar{e}} (D^{\mu}\phi)^{a}{\phi^\dagger}^{\bar{a}} \Sigma^{\bar{b}b} \Sigma^{\bar{c}c} \Sigma^{\bar{d}d} \Sigma^{\bar{e}e}$ where fields with indices with a bar transform with respect to the anti-fundamental representation. Also, similar operators contribute to the spectrum of the scalar channel. Performing the FMS prescription, we obtain that the leading order contribution comes from the elementary massless photon and the next-level state has the mass of the $Z$ boson in both cases. Thus, the gauge-invariant spectrum consists of two massless vector fields which are not observed in nature in case the theory exhibits a discrete $\mathbb{Z}_{2}$ symmetry. 

At first glance, these problems can be solved by introducing a $\mathbb{Z}_{2}$ breaking term on the level of the Lagrangian. Then, we can classify the states only according to the global $\U(1)$ quantum number and the spectrum would contain only one low lying scalar with the mass of the lightest elementary Higgs as well as one massless vector particle with a possible resonance which could be interpreted as the $Z$ boson. However, there is no obvious way to construct gauge-invariant composite states that expand to only one elementary $W^{\pm}$ field.

It is possible to construct states with open $\U(1)$ quantum number according to Eq.~\eqref{eq:fund-U(1)vec}. However, these are generically to heavy as they contain at least three leptoquarks (as well as an elementary $W$) at leading order and thus, their mass is of the order of the GUT scale. Of course, there are many other potential operators within this channel, e.g., $\epsilon^{abcde}\phi^{a}(\Sigma\phi)^{b}(\Sigma^{2}\phi)^{c}(\Sigma^{3}\phi)^{d}(D^{\mu}\phi)^{e}$ which contain only a single gauge field. Unfortunately, the leading order contribution in the fluctuating fields vanishes identically due to the antisymmetric properties of the epsilon tensor if the QCD vacuum shall be unbroken. Therefore, this state will generically be heavier than the mass of the lightest elementary Higgs. As we only have the fundamental vectors $\phi$ and $\Sigma^{n}\phi$ which do not contain an elementary gauge field, we do not see any comparatively simple operator which can expand to an object with the desired properties.

Thus, the number and masses of the gauge-invariant states are actually incompatible with the standard model, as the low-lying charged gauge bosons, the $W^{\pm}$, are not appropriately reproduced. 
Moreover, the FMS analysis demonstrates that $\mathbb{Z}_{2}$ violating interactions are needed to avoid a doubling of the photon and light Higgs state, which are only distinguished by their global $\mathbb{Z}_{2}$ parity, in contrast to the elementary spectrum. 
For these reasons, the bosonic sector of $\SU(5)$ is likely not an legitimate extension of the bosonic sector of the standard model already from a purely field-theoretical point of view.

\section{Conclusions}

In conclusion, the FMS mechanism was used to investigate the differences of the physical spectrum and the spectrum of the elementary fields for a number of prototype theories, but also for the bosonic sector of a more realistic GUT. In this course, we derived predictions for the physical spectra, which in almost all cases differed from the expectations based on the elementary fields. These differences are almost always qualitative, and can therefore not be expected to be removed by radiative corrections. In particular, based on these arguments the usual construction of $\SU(5)$ as an extension of the standard model would be ruled out already on structural grounds as a possible candidate for a grand-unified theory, even if it would not be ruled out for quantitative reasons \cite{Bohm:2001yx,Georgi:1974yf,Nishino:2009aa}.
Besides the actual particle spectrum, our investigations demonstrate that, for instance, the computation of a possible proton decay has to indispensably be rethought. 
In addition a variety of other aspects with cosmological implications might change. 

The predictions for the gauge-invariant physical states can be checked using non-perturbative methods. Of course, once the mechanism has been established, analytic predictions for other theories can be made with similar ease, and confidence, as in ordinary perturbation theory. It is worth emphasizing, that each and every non-perturbative test using lattice calculations of the mechanism and of concrete predictions agree with the analytical results, especially those presented here \cite{Shrock:1985un,Shrock:1985ur,Lee:1985yi,Maas:2012tj,Maas:2013aia,Maas:2016ngo,Maas:unpublishedtoerek}. This is even true if the predictions are in qualitative disagreement to ordinary perturbation theory \cite{Lee:1985yi,Maas:2016ngo,Maas:unpublishedtoerek}. These highly non-trivial tests give confidence in the methods and the multitude of further predictions presented here.

A next logical step, besides continuing checks on the lattice, would be to investigate current candidates for BSM physics, including also the fermionic sector along the lines of \cite{Frohlich:1980gj,Frohlich:1981yi,Egger:2017tkd,Torek:2018qet}, to see whether conflicts arise for some of them as well. This does not need to be the case, as the explicit example of 2HDMs shows \cite{Maas:2016qpu}, but may happen as have been seen here. It would also be good to further develop the tools designed here to allow a quicker assessment of which theories may harbor conflicts, and which not.

\section*{Acknowledgments}

It is a pleasure to thank A.\ Wipf and L.\ Pedro for valuable discussions and for a critical reading and useful comments on the manuscript. RS thanks the Carl-Zeiss Stiftung for support through a postdoc fellowship. PT has been supported by the FWF doctoral school W1203-N16.

\appendix

\section{The structure of the state space}\label{a:state}

When considering the adjoint case in Section \ref{sec:adjoint} an additional complication arose.

This complication arose in the following way. In the fundamental case, every gauge orbit is either only zero field, or belongs to an orbit with symmetry group $\SU(N-1)$, as always a gauge transformation exists which rotates a given scalar field locally into a vector with only a single-non-zero component. These two distinct classes are called strata, and the corresponding symmetry groups, $\SU(N)$ for zero scalar field and $\SU(N-1)$ otherwise, are the little groups \cite{O'Raifeartaigh:1986vq}. Thus, in this case there is only a single special orbit, the vacuum, and all others behave in the same way.

When moving outside this case, the situation becomes more complex. In the following the SU(3) case with an adjoint scalar field will be used as an example, and the most general case has not yet been solved in a constructive way, to the knowledge of the authors. 
Note, however, that the ranks of the little groups, and thus the size of the Cartans, play an important role in determining the number of different little groups \cite{O'Raifeartaigh:1986vq}.

In the general case, there are not only two strata and little groups, but more. For instance, for the SU(3) case with adjoint Higgs there are three: SU(3), SU(2)$\times$U(1), and U(1)$\times$U(1). Any value of the Higgs field can have only one of these little groups as invariance groups, and the set of all such orbits is again the corresponding stratum of the little group. Thus, there is no gauge transformation moving a value of the Higgs field from one stratum to another, and belonging to a stratum is a gauge-independent statement. Thus, corresponding gauge-invariant quantities to state this fact exists, these are merely the invariant polynomials of the group and the representation \cite{O'Raifeartaigh:1986vq}. For instance, \eqref{eq:sigma3} yields to which stratum a field locally belongs for the $\SU(3)$ case.

There is now a twist to this group-theoretical problem in a field theory. The distinction is local. The scalar field is a field, and its value changes from point to point. Especially, a scalar field can belong to any stratum at different space-time points. Thus, the distinction is not meaningful in a global way. Still, because a gauge transformation acts on the Higgs field locally, this feature is again locally gauge-invariant. Thus, the function \eqref{eq:sigma3} locally characterizes in the example the stratum of the scalar field.

Likewise, it is possible to characterize the space-time average of any scalar field configuration by the stratum to which it belongs. This will be invariant under global gauge transformations.

The question is now how this affects the main part of the text.

First of all, this does not affect perturbation theory. Because perturbation theory is a small field expansion, perturbation theory will stay inside a given stratum, by definition, as the characterization in terms of invariants is discontinuous \cite{O'Raifeartaigh:1986vq}.

However, this is a problem when attempting to fix the gauge \eqref{eq:gaugeAdj} beyond perturbation theory. If the vector $\Sigma^{0}_{i}$ belongs to a given stratum, and the Higgs field at a point $x$ belongs to a different stratum, then the term $[\Sigma^{0},\Sigma]$ vanishes at this point.\footnote{Note that the global $\mathbb{Z}_2$ symmetry, if not explicitly broken, is broken by the gauge condition, yielding a diagonal $\mathbb{Z}_2$ subgroup.} Thus, at this point the gauge condition degenerates to the covariant gauge condition. However, the gauge was chosen such that the gauge condition rotates the space-time average of the Higgs field into the direction of $\Sigma^{0}$, which is part of a fixed stratum. This is impossible, as noted above, if the average of a gauge orbit belongs to a different stratum. Thus, the gauge condition is not fulfilled on this gauge orbit. Instead, the orientation within the stratum is not affected, and thus ultimately still an average over the directions inside the stratum is performed, yielding again a zero expectation value for this gauge orbit. Hence, on such a gauge orbit the gauge condition \eqref{eq:gaugeAdj} degenerates into the covariant gauge condition.

The path integral therefore decomposes into a sum of distinct parts. One contains the orbits for which the gauge condition in terms of the space-time average can be fulfilled, and the remainder contains the strata where this is not the case, and the vacuum expectation value vanishes. Consequently, the vacuum expectation value will still be the desired one, as the second part does not contribute, provided the measure of gauge orbits in the first part is not of size zero.

Unfortunately, this implies also that it is not possible to distinguish between strata using expectation values. Though, e.g., \eqref{eq:sigma3} is gauge-invariant, its actual value is determined by weighting the value for every gauge orbit by the exponentiated action, and averaging over the orbits of the different strata. Its expectation value is therefore possibly continuous throughout the phase diagram of the theory.\footnote{This is not necessarily so. But there is always also the QCD-like phase, which technically belongs to the full group, as no direction is preferred. However, the corresponding stratum has measure zero, it is only the vacuum, and it can thus not arise by any other means than cancellation.}

For the calculation of the spectrum using the FMS mechanism this has the following consequence. Given the arguments above, the vacuum expectation has still the same direction. However, fields belonging to a different stratum do not have a small fluctuation around this vacuum expectation value. Thus, for correlators holds, symbolically,
\begin{align}
\langle &O^\dagger(x) O(y)\rangle = \nonumber\\
&=\int{\cal D}A\left(~\int_\text{Selected stratum}{\cal D}\phi~O^\dagger(x)O(y)~e^{iS}\right.\nonumber\\
&\left.+\int_\text{Other strata}{\cal D}\phi~O^\dagger(x)O(y)~e^{iS}~\right)\nonumber\\
&=\langle O^\dagger(x)O(y)\rangle_e+\langle O^\dagger(x) O(y)\rangle_n\nonumber\;,
\end{align}
\noindent i.e., it decomposes into two correlators, of which one is meaningfully expendable (index $e$) around the vacuum expectation value, while the other is not. The latter can, in principle, have any arbitrary pole structure, since no meaningful perturbative expansion is possible.

For the purpose of the main text there are two possible sets of assumptions for proceeding:
\begin{itemize}
 \item There are no non-trivial (non-scattering) pole structures in the second correlator, and thus the pole structure of the first term, determined using the FMS expansion, completely describes the physics. This does not imply that if the physical spectrum differs for different choices of expansion strata this gives rise to a physical distinction. The different results are not changing the multiplicity, and the change can come about by gradual evolution.
 \item The second correlator harbors the pole structures which would be obtained when fixing the gauge based on the other stratas. Then the physical spectrum would be obtained by the union of all spectra obtained by using the FMS mechanism around every possible vacuum expectation value.
\end{itemize}
At the current time we do not have arguments in favor of either possibility. It is quite possible that this is dynamically decided. Note, e.g., that if the global $\mathbb{Z}_2$ parity is explicitly broken, the absolute minimum always favors the maximal little group for $\SU(3)$ \cite{O'Raifeartaigh:1986vq}. Thus, it is entirely possible that if the potential spontaneously breaks this global symmetry the first case may be appropriate, and then only one of the strata contributes to the spectrum, and otherwise the second. This will require a full non-perturbative investigation of the spectrum and the phase diagram to understand better. In the main text we therefore discuss all possible spectra.

\section{Analysis of open U(1) states}\label{sec:full}
In this appendix, we sketch the computation of the bound state spectrum regarding operators with open $\U(1)$ quantum numbers for a Higgs field in the fundamental representation. We again choose the convenient gauge in which $n^{a} = \delta^{aN}$. In order to keep the computation transparent, we introduce the following abbreviations for the gauge fields, 
\begin{align}
 A^{\mu} = A^{\mu}_{i} T_{i} \equiv 
 \begin{pmatrix} \bold{\tilde{A}^{\mu}_{0}} & X^{+\mu} \\ X^{-\mu} & {Z'}^{\mu} \end{pmatrix}.
\end{align}
The matrix $\bold{\tilde{A}^{\mu}_{0}}$ is the $(N-1)\times (N-1)$ submatrix containing the $(N-1)^{2}-1$ massless gauge fields, $\bold{\tilde{A}^{\mu}_{0}} = A^{\mu}_{i}\tilde{T}_{i}$ with $\tilde{T}$ the generators of the $\SU(N-1)$ subgroup. The abbreviation $X^{+\mu}$ contains the gauge fields in the $N$th column of $A^{\mu}$ except for the $N$th element, thus forming an $N-1$ component complex column vector
\begin{align}
 X^{+\mu} \equiv \begin{pmatrix} X^{+\mu,1} \\ \vdots \\X^{+\mu,N-1}  \end{pmatrix}  
 = \frac{1}{2} \begin{pmatrix} A^{\mu}_{(N-1)^{2}} - \I\, A^{\mu}_{(N-1)^{2}+1} \\ \vdots \\ A^{\mu}_{N^{2}-3} - \I\, A^{\mu}_{N^{2}-2}  \end{pmatrix},
\end{align}
and $X^{-\mu} = (X^{+\mu})^{\dagger}$. The fields $X^{+\mu}$ and $X^{-\mu}$ encode the $2(N-1)$ degenerated massive gauge bosons with mass $\mA$. Finally, ${Z'}^{\mu}$ is a short cut for the $N\times N$ element of $A^{\mu}$ given by ${Z'}^{\mu} = - \frac{1}{2}\sqrt{2/N} A^{\mu}_{N^{2}-1}$ and encodes the heaviest gauge boson with mass $\MA$.

With this reformulation of the gauge boson matrix, it is straightforward to analyze the spectrum of the operators in the open $\U(1)$ channel. Starting with the vector operator \eqref{eq:fund-U(1)vec}, we obtain in leading order in the FMS expansion,
\begin{align}
  O^\mu_{1^-_1}  &= \I g \left( \frac{v}{\sqrt{2}} \right)^{N} \epsilon^{a_{1}\cdots a_{N}}\, n^{a_{1}}(A_{\nu_{1}}n)^{a_2} (F^{\nu_{1}}_{\phantom{\nu_1}\nu_{2}}n)^{a_{3}} \cdots \notag \\
  &\qquad\quad \cdots (F^{\nu_{N-2}\mu} \, n)^{a_N} + \mathcal{O}(\varphi).
  \label{eq:openU1-exp}
\end{align}
In order to obtain the elementary field content of the gauge invariant operator, we have to compute the components of the $\SU(N)$-vectors $A^{\mu}n$ and $F^{\mu\nu}n$. They read,
\begin{align}
 A^{\mu}n = \begin{pmatrix} X^{+\mu} \\ {Z'}^{\mu} \end{pmatrix},
\end{align}
and 
\begin{align}
 &F^{\mu\nu} n \notag\\
 &= \begin{pmatrix} \pd^{\mu} X^{+\nu} - \pd^{\nu} X^{+\mu} \\  \pd^{\mu}{Z'}^{\nu} - \pd^{\nu}{Z'}^{\mu} \end{pmatrix} 
 + 2\I g \begin{pmatrix} \bold{\tilde{A}_{0}}^{[\mu} X^{+\nu]}_{\phantom{0}} + {X^{+}}^{[\mu} {Z'}^{\nu]} \\ X^{-[\mu} X^{+\nu]}  \end{pmatrix}.
 \label{eq:Fn}
\end{align}
The right-hand side of Eq.~\eqref{eq:openU1-exp} is nonvanishing, only if the $(N-1)$-tuple $(a_{2},a_{3},\cdots,a_{N})$ is given by $(1,2,\cdots,N-1)$ or a permutation of these numbers due to the antisymmetry of the epsilon tensor and $n^{a_{1}} = \delta^{Na_{1}}$. Neglecting for a moment the contribution from the commutator of $F^{\mu\nu}$, $O_{1^{-}_{1}}$ contains schematically the product $X^{+ 1}X^{+ 2}\cdots X^{+ N-1}$ of the $(N-1)$ different fields stored in $X^{+}$ at leading order in the FMS expansion. 

Computing the propagator and employing the simple constituent model, we predict at tree level that the mass of the operator \eqref{eq:fund-U(1)vec} is given as the sum of the masses of the $N-1$ elementary massive gauge fields, $m_{1^{-}_{1}} = (N-1)\mA$. 

Moreover, the investigated operator contains an excitation at mass $m_{1^{-}_{1}}^{*} = (N-1)\mA+\MA$. This can easily be seen from the last term in Eq.~\eqref{eq:Fn} coming from the commutator of the two gauge fields in $F^{\mu\nu}$. 
From this additional term, we can read of that the correlator of the gauge-invariant bound state operator also contains at next-to-leading order in the gauge fields the propagation of the different $N-1$ fields $X^{+\mu}$ and an elementary gauge boson which is either one of the massless gauge fields stored in $\bold{\tilde{A}_0^\mu}$ or the heaviest gauge boson described by $Z'$. 
For $\MA<\mh$, the latter predicts the first next-level state of $O_{1^-_1}$. 
In case $\MA>\mh$, this state will be either a trivial scattering state of the ground state with the bound state of the scalar singlet or a resonance which can decay to the ground state and the bound state of the scalar singlet, $(1^{-}_{1})^{*} \to 1^{-}_{1} + 0^{+}_{0}$. 

As the operator $O_{1^-_1}$ is build from $F^{\mu\nu}n$ precisely $N-2$ times, we get similarly excited states with mass $(N-1)\mA + 2 \MA$, $\cdots$, $(N-1)\mA + (N-2) \MA$ as we have schematically $X^{+}(X^{+}+\bold{\tilde{A}_{0}}X^{+} + X^{+}Z')^{N-2}$. These are trivial scattering states (or might be resonances) of the ground state (or its first excitation) and the vector or scalar singlets regarding the $\U(1)$ independent of the relation between $\mh$ and $\MA$. For instance the state with mass $(N-1)\mA + 2\MA$ can be viewed as a scattering state of the ground state of $O_{1^{-}_1}$ and two ground states of $O_{1^{-}_0}$.

In addition, the next-to-leading order contribution in the FMS expansion $\sim\!\varphi$ contributes also to the spectrum with mass $(N-1)\mA+\mh$ but is likely to be always a trivial scattering state of the ground state of $O_{1^-_1}$ and $O_{0^+_0}$ within our first order approximation.

In full analogy, the ground state spectrum as well as the higher excitations of the scalar operator with open $\U(1)$ quantum number $O_{0^{+}_{1}}$, see Eq.~\eqref{eq:fund-U(1)sca}, can be derived. The ground state has mass $(N-1)\mA$ and possible additional particles might be encoded in the next-level state with mass $(N-1)\mA + \MA$ for $\MA<\mh$.

\section{SU(3) with two Higgs in the fundamental representation}\label{sec:su32}
We sketch the spectrum of a gauge theory with two Higgs fields in the fundamental representation in the following. 
For simplicity, we only discuss the case where the gauge group is $\SU(3)$ but generalize the results of \cite{Torek:2015ssa} where a similar investigation was performed. The Lagrangian which we use is of the form
\begin{align}
\mathcal{L} = -\frac{1}{4}F_{i~\mu\nu} F^{\mu\nu}_i + 
\sum_{\alpha=1}^2(D_\mu \phi_{\alpha})^\dagger(D^\mu \phi_{\alpha}) - V(\phi_1,\phi_2),
\label{eq:L2H}
\end{align}
where $\alpha$ labels the different Higgs flavors. Depending on the precise form of the scalar potential, the theory can have different global symmetries, e.g., a global $\U(1) \times \U(1)$ symmetry for potentials of the form $V(\phi^{\dagger}_1\phi_1,\phi^{\dagger}_2\phi_2)$ or a global $\SU(2)\times\U(1)$ symmetry for $V(|\phi_1|^2+|\phi_2|^2)$.
We will focus to this end on the latter case, where the potential has the largest custodial symmetry. 
This symmetry can be used to construct gauge-invariant states which have distinct $\SU(2)\times \U(1)$ transformation properties. As usual, we start our discussion by summarizing the gauge-variant spectrum of the theory. 

Depending on the alignment of the vevs different breaking patterns exist. If the two vevs are (anti-)parallel the breaking pattern reads $\SU(3)\rightarrow \SU(2)$. In all other cases the breaking pattern is $\SU(3)\rightarrow \SU(2)\rightarrow 1$. The elementary spectrum of the former case is listed in the left panel of Tab.~\ref{tab:su32_par}.  In Tab.~\ref{tab:su32_orth} we restrict the discussion on vevs which are orthogonal. In case the second vev has a parallel as well as an orthogonal component can also straightforwardly be computed but leads only to further mass splits in the elementary spectrum and does not give new insights in the structures of the gauge-invariant spectrum.

\begin{table}[t!]
\begin{center}
\begin{tabular}{l|llc|lllc}

\multicolumn{4}{r}{\qquad elementary spectrum} & \multicolumn{4}{c}{\quad gauge-invariant spectrum \quad} \cr
\toprule
$J^P$ & Field & Mass & Deg & $(I,I^3)$ & Operator & Mass  & Deg\cr
\hline

$0^+$ & $\varphi_1^5$ & $\mh$ & $1$ & $(0,0)$ & $O_{0^+_0}$ & $\mh$ & $1$\cr
 & $\varphi_2^{1,\dots,6}$ & $0$ & $6$ & $(1,I_3)$ & $O_{0^+_{1,2,3}}$ & $0$ & $3$ \cr
\hline
$1^-$ & $A^\mu_{8}$ & $\MA$ & $1$  & $(0,0)$ & $O^{\mu}_{1^-_0}$ & $\MA$ & $1$\cr
 & $A^\mu_{4,\dots,7}$ & $\mA$ & $4$ & $(1,I_3)$ & $O^{\mu}_{1^-_{1,2,3}}$ & $\MA$ & $3$\cr
 & $A^\mu_{1,2,3}$ & $0$ & $3$ &  &  &  & \cr

\toprule
\end{tabular}
\caption{Summary table for the case with two parallel vevs. Here, we used $n_\alpha^a = \delta^{a,3}$, $\alpha=1,2$. \\
Left: Gauge-variant spectrum of an $\SU(3)$ gauge theory with two scalar fields in the fundamental representation and $\SU(2)\times \U(1)$ custodial symmetry. The fields listed here are all mass-eigenstates. \\
Right: Gauge-invariant (physical) spectrum of the theory to leading order. \\
$\mh$ denotes the mass of the  massive Higgs, $\MA$ is the mass of the heaviest gauge boson and $\mA$ the mass of the degenerate lighter gauge bosons. $(I,I^3)$ are the quantum numbers of the  global symmetry group $\SU(2)$. We only consider $\U(1)$-singlet states here.}
\label{tab:su32_par}
\end{center}
\end{table}

\begin{table}[t!]
\begin{center}

\begin{tabular}{l|llc|lllc}

\multicolumn{4}{r}{\qquad elementary spectrum} & \multicolumn{4}{c}{\quad gauge-invariant spectrum \quad} \cr
\toprule
$J^P$ & Field & Mass & Deg & $(I,I^3)$ & Operator & Mass  & Deg\cr
\hline

$0^+$ & $\varphi_1^5$ & $\mh$ & $1$ & $(0,0)$ & $O_{0^+_{0}}$ & $\mh$ & $1$\cr
 & $\varphi_2^{1,5,6}$ & $0$ & $3$ &  $(1,I_3)$ & $O_{0^+_{1,2,3}}$ & $0$ & $3$ \cr
\hline
$1^-$ & $A^\mu_{8}$ & $\MA$ & $1$  & $(0,0)$ & $O^{\mu}_{1^-_0}$ & $\mA$ & $1$\cr
 & $A^\mu_{6,7}$ & $m_{\mathrm{A}_{6,7}}$ & $2$ &  $(1,\pm 1)$ & $O^{\mu}_{1^-_{1,2}}$ & $m_{\mathrm{A}_{4,5}}$ & $2$\cr
 & $A^\mu_{4,5}$ & $m_{\mathrm{A}_{4,5}}$ & $2$ & $(1,0)$ & $O^{\mu}_{1^-_3}$ & $\mA$ & $1$\cr
 & $A^\mu_{3}$ & $\mA$ & $1$ &  &  &  & \cr
 & $A^\mu_{1,2}$ & $m_{\mathrm{A}_{1,2}}$ & $2$ &  &  &  & \cr

\toprule
\end{tabular}
\caption{Summary table for the case with orthogonal vevs. Here, we use the special choice $n_1^a = \delta^{a,3}$ and $n_2^a = \delta^{a,1}$. \\
Left: Gauge-variant spectrum of an $\SU(3)$ gauge theory with two scalar fields in the fundamental representation and $\SU(2)\times \U(1)$ custodial symmetry. The fields listed here are all mass-eigenstates. \\
Right: Gauge-invariant (physical) spectrum of the theory to leading order. 
$\mh$ denotes the mass of the Higgs and the masses of the gauge bosons have the following ordering: $\mA<m_{\mathrm{A}_{1,2}}<m_{\mathrm{A}_{6,7}}<m_{\mathrm{A}_{4,5}}<\MA$. $(I,I^3)$ are the quantum numbers of the global symmetry group $\SU(2)$. We only consider $\U(1)$-singlet states here.}
\label{tab:su32_orth}
\end{center}
\end{table}

Having settled the perturbative spectrum we can now focus on the gauge-invariant (physical) spectrum of the theory by applying the FMS mechanism.  We start with operators which are $\U(1)$ singlets. Due to the $\SU(2)$ custodial symmetry we can arrange our operators in singlets and triplets for $J^P=0^+,1^-$:
\begin{align}
\begin{split}
&O_{0^+_{\hat{i}}} = \phi_\alpha^\dagger~\tau^{\alpha \beta}_{\hat{i}}~\phi_\beta, \qquad 
O_{1^-_{\hat{i}}}^{\mu} = \phi_\alpha^\dagger~\tau^{\alpha \beta}_{\hat{i}}~D_\mu~\phi_\beta, \\
&\tau_{\hat{i}}\in\left\{\mathbbm{1},\tau_+=\frac{\sigma_1+i\sigma_2}{2},\tau_-=\frac{\sigma_1-i\sigma_2}{2},\sigma_3\right\}
\end{split}
\label{eq:su32ops}
\end{align}
with $\hat{i}=0,1,2,3$ and $\sigma_{1,2,3}$ are the usual Pauli matrices. The operators for the case $\hat{i}=0$ are the singlets and $\hat{i}=1,2,3$ are the triplet operators for both $J^P$-$\U(1)$-singlet channels.

Using again the FMS mechanism to leading order, we obtain the results listed in the right panels of Tab.~\ref{tab:su32_par} and \ref{tab:su32_orth}.

We expect in the $0^+$ channel for both, parallel and orthogonal vevs, one massive state with the mass of the perturbative Higgs in the $\SU(2)$-singlet channel and in the triplet we expect three degenerate massless states. In the $1^-$ channel there are four degenerate massive states with the mass of the heaviest gauge boson in the case of parallel vevs. In the case of orthogonal vevs there is one state in the $\SU(2)$-singlet channel with the mass of the lightest gauge boson which also appears in the triplet channel. In this channel there are also two degenerate states with masses of the second heaviest gauge boson(s). 
In both cases this in contradiction to the elementary spectrum (cf. left panels of Tab.~\ref{tab:su32_par} and \ref{tab:su32_orth}).

There is one interesting observation: If the vevs are parallel then the triplets in the $0^+$ and in the $1^-$ channel are degenerate, which one would expect due to the Wigner-Eckart theorem. However, if the vevs are orthogonal then the degeneracy splits (at least in the $1^-$ channel). That appears due to the fact that in the case of parallel vevs there is still a $\SU(2)$ rotation left which can be applied to two of the three components of the vevs. This is not the case if the vevs are orthogonal.

Of course we can construct, in analogy to Sec.~\ref{sec:fundamental}, objects similar to baryons, i.e., states with open $\U(1)$ quantum numbers. An example set of these states are ($\alpha=1,2$):
\begin{align*}
\begin{split}
O_{0^+_1,\alpha} &= \epsilon^{abc}\phi_1^a \phi_2^b (D_\mu D^\mu \phi_\alpha)^c,\quad O_{0^+_{-1},\alpha} = (O_{0^+_1,\alpha})^\dagger, \\
O^{\mu}_{1^-_1,\alpha} &= \epsilon^{abc}\phi_1^a \phi_2^b (D^\mu \phi_\alpha)^c, \qquad\,\, O^{\mu}_{1^-_{-1},\alpha} = (O^{\mu}_{1^{-}_{1},\alpha})^\dagger.
\end{split}
\end{align*}

It is straightforward to see that if the vevs are parallel, the leading order contribution of the FMS expansion vanishes due to the antisymmetry of the $\epsilon$-tensor.  Therefore, we discuss only on the case where the vevs are orthogonal. For simplicity we restrict to the case $v_1\gg v_2$ as it is the typical situation in GUTs. Then $\mA=m_{\mathrm{A}_{1,2}}$ and  $m_{\mathrm{A}_{4,5}}=m_{\mathrm{A}_{6,7}}\equiv \mA^\prime$. After expanding the corresponding correlators to leading order in the FMS expansion and in perturbation theory (a similar strategy as in App.~\ref{sec:full}) we obtain the following spectrum of  states with an open $U(1)$ quantum number for the theory in the above limit:
\begin{itemize}
\item $0^+$ channel: Ground state mass of $2\mA$ and excited states $\mA+\mA^\prime < 2\mA^\prime < \mA+\MA<\mA^\prime + \MA$.
\item $1^-$ channel: Ground state mass of $\mA$ and an excited state with a mass of $\mA^\prime$.
\end{itemize}
Note that all these states have conjugated partners, describing the particle and anti-particle, respectively.

\bibliography{bib}

\end{document}